\documentclass[aos]{imsart}

\RequirePackage{amsthm,amsmath,amsfonts,amssymb}
\RequirePackage[authoryear]{natbib}
\RequirePackage[colorlinks,citecolor=blue,urlcolor=blue]{hyperref}
\RequirePackage{graphicx}
\RequirePackage{bm}
\RequirePackage{enumerate}
\RequirePackage{url}
\RequirePackage{algorithm}
\RequirePackage{algpseudocode}
\RequirePackage{float}
\RequirePackage{multirow}
\RequirePackage{subfigure}
\RequirePackage{mathabx}

\startlocaldefs
\theoremstyle{plain}
\newtheorem{theorem}{Theorem}[section]
\newtheorem{lemma}[theorem]{Lemma}
\newtheorem{corollary}[theorem]{Corollary}
\theoremstyle{remark}

\newtheorem{assumption}{Assumption}

\newcommand{\leftsecond}{\left\{}
\newcommand{\rightsecond}{\right\}}
\newcommand{\leftthird}{\left[}
\newcommand{\rightthird}{\right]}
\newcommand{\leftdot}{\left.}
\newcommand{\rightdot}{\right.}

\newcommand{\oE}{\operatorname{E}}
\newcommand{\coti}{i=1,\ldots,K}
\newcommand{\cotj}{j=1,\ldots,n}
\newcommand{\sumi}{\sum_{i=1}^K}
\newcommand{\sumj}{\sum_{j=1}^n}
\newcommand{\iterO}{{(t)}}
\newcommand{\iterN}{{(t+1)}}

\endlocaldefs

\begin{document}

\begin{frontmatter}
\title{Empirical Likelihood Inference over Decentralized Networks}
\runtitle{Network Empirical Likelihood}

\begin{aug}

\author{\fnms{JINYE}~\snm{DU}\ead[label=e1]{dujinye14@mails.ucas.ac.cn}}
\and
\author{\fnms{QIHUA}~\snm{WANG}\ead[label=e2]{qhwang@amss.ac.cn}}
\address{Academy of Mathematics and Systems Science, Chinese Academy of Sciences,}
\address{University of Chinese Academy of Sciences\printead[presep={,\ }]{e1,e2}}

\end{aug}

\begin{abstract}
As a nonparametric statistical inference approach, empirical likelihood has been found very useful in numerous occasions.
However, it encounters serious computational challenges when applied directly to the modern massive dataset.
This article studies empirical likelihood inference over decentralized distributed networks, where the data are locally collected and stored by different nodes.
To fully utilize the data, this article fuses Lagrange multipliers calculated in different nodes by employing a penalization technique.
The proposed distributed empirical log-likelihood ratio statistic with Lagrange multipliers solved by the penalized function is asymptotically standard chi-squared under regular conditions even for a divergent machine number.
Nevertheless, the optimization problem with the fused penalty is still hard to solve in the decentralized distributed network.
To address the problem, two alternating direction method of multipliers (ADMM) based algorithms are proposed, which both have simple node-based implementation schemes.
Theoretically, this article establishes convergence properties for proposed algorithms, and further proves the linear convergence of the second algorithm in some specific network structures.
The proposed methods are evaluated by numerical simulations and illustrated with analyses of census income and Ford gobike datasets.
\end{abstract}



\end{frontmatter}


\section{Introduction}

The rapid advancement of modern technology has facilitated the collection of unprecedentedly large datasets, which poses tremendous computation and storage challenges to statistical analysis.
Given an exceedingly large dataset, a single machine may lack the necessary memory for storage and processing.
Moreover, in some practical scenarios, data is collected locally and remains distributed across multiple local machines, such as mobile devices, local governments, research labs, hospitals, and wireless sensors \citep{jochems2016distributed,guo2020adaptive}.
Furthermore, aside from the memory limitation of a single machine, it may be impractical to transfer all the data to a single machine due to limitation of the network bandwidth, privacy, security or ethical concern.
To fully utilize the data residing in multiple machines, distributed statistical analysis have drawn plenty of attentions in various areas, including M-estimation \citep{shi2018massive,jordan2019communication,fan2023communication}, high-dimensional test and estimation \citep{lee2017communication,battey2018distributed,hector2021distributed}, quantile regression \citep{chen2019quantile,volgushev2019distributed}, principal component analysis \citep{fan2019distributed,huang2021communication,chen2022distributed}, support vector machine \citep{lian2018divide,wang2019distributed}, and so on.

The main strategy of the distributed statistical approaches mentioned above is to calculate local statistics based on the dataset in local machines, and aggregate the local statistics into a summary statistic.
However, such distributed approaches heavily rely on the central machine to communicate with every local machines for the purpose of aggregating the local statistics.
Such a type of architecture suffers from several serious limitations including privacy protection, network stability and bandwidth requirement \citep{wu2022network}, especially when the number of local machines diverges.
To address these problems, researchers advocate the idea of decentralized distributed computation \citep{vanhaesebrouck2017decentralized,tang2018decentralized}.
The key feature is that there is no central machine involved for computation, and all the computation-related communications should be limited between directly connected nodes.
By mutually sharing the local statistics, all the nodes in the decentralized network collaborate with their neighbors to perform the global statistical analysis.
In recent years, driven by practical considerations, there has been an increasing interest in the statistical properties of decentralized distributed approaches, including M-estimation \citep{wu2022network}, heterogeneity analysis \citep{zhang2022learning}, and so on.
The existing literature primarily focuses on the point estimating problems over decentralized networks.  Unfortunately, they cannot be applied to statistical inference because either the asymptotic distributions of the estimates 
are not derived or the asymptotic variance estimates cannot be computed over the decentralized network.

Compare to  the normal approximation based inference,  empirical likelihood \citep{owen1988empirical,owen1990empirical} has many advantages. It is demonstrated effective in scenarios with less restrictive distributional assumptions for statistical inferences.
It has many nice properties: automatic determination of the shape of confidence regions \citep{owen1988empirical}, incorporation of auxiliary information through constraints \citep{qin1994empirical}, self-studentization without calculating the covariance matrices, range respecting and transformation preserving \citep{owen2001empirical}.
For these reasons, empirical likelihood has found many applications in many fields such as meta-analysis \citep{qin2015using,huang2016efficient,han2019empirical,zhang2020generalized}, high-dimensional data analysis \citep{tang2010penalized,lahiri2012penalized,leng2012penalized,chang2018new,chang2021high}, censored data analysis \citep{li2003empirical,he2016empirical,tang2020penalized}, missing data analysis \citep{wang2002missing,wang2009empirical}, measurement error analysis \citep{wang2002error}, and so on.
For a more comprehensive review, see \cite{owen2001empirical}, \cite{chen2009review}, and \cite{lazar2021review}.
This motivates us to develop a distributed empirical likelihood method over the decentralized network.

It is noted that distributed empirical likelihood inference has been developed by \cite{zhou2023distributed}.
However, the existing work assumes the presence of a central machine connecting to all the local machines, which cannot be extended to the case of decentralized distributed network.
In this article, we study empirical likelihood inference for estimating equation problems over decentralized distributed network, and propose specific decentralized distributed algorithms with node-based structures.
In this situation, computations are carried out in each node, and communications are limited between directly connected nodes.
%
The main difficulty and challenge in developing empirical likelihood inference over decentralized distributed network is to aggregate information from datasets in different nodes, and calculate the global Lagrange multiplier that has no explicit expression.
This is quite challenging.
The existing methods are unavailable to solve it over a decentralized networks.
To address it, instead of calculating the Lagrange multiplier by solving the corresponding equation, we compute it by solving an optimization problem defined by the convex dual representation.
Moreover, we introduce an equivalent representation adjusted to the pre-specified communication-based network structure for the original optimization problem.
The equivalent representation is a constrained optimization problem with network structure-based constraints.
Further, we employ a penalization technique to solve the conditional optimization problem and transform it to an unconstrained optimization problem, where the fused penalty term indicates the neighbour relationship between nodes.
Theoretically, we demonstrate the equivalence property between the solutions of the unconstrained optimization problem and the original equation, which ensures that the distributed empirical log-likelihood ratio statistic with Lagrange multipliers calculated by the unconstrained optimization problem is asymptotically standard chi-squared.

Nevertheless, the unconstrained optimization problem is still hard to solve in the decentralized distributed network because information exchanges are limited between the nodes which are directly connected.
In this article, we propose two algorithms to solve the unconstrained optimization problem.
Firstly, motivated by \cite{hallac2015network}, we develop an ADMM-based algorithm called Pairwise Copy Method (PCM) and establish its convergence property.
Although PCM has clear intuition and employs a novel decoupling technique, its convergence rate is hindered by an excessive number of parameters.
Meanwhile, the update strategy of PCM relies on time-consuming numerical computation, which is unnecessary in the framework of ADMM.
To fix the imperfection of PCM, we design another algorithm called Modified Approximation Objective Method (MAOM).
Comparing with PCM, MAOM only introduces half of the parameters in the optimization procedure, and employs adaptive second-order approximations in each iteration.
In each iteration, the adaptive approximation provides closed-form solutions.
This greatly reduces the computational load and accelerating computation speed.
To enhance the convexity of the approximations, we also add a strict convex quadratic regularization function to the approximated objective function, which can be regarded as the proximal function in the sense of a matrix norm.
After meticulously designing the regularization function, the algorithm has a simple node-based implementation scheme.
Theoretically, we prove the convergence property of MAOM, and further establish its linear convergence rate in some specific network structures that can be generated by arbitrary connected networks.

The paper is organized as follows.
In Section \ref{sec:framework}, we develop empirical likelihood inference for estimating equation problems in the framework of the decentralized distributed network, and propose a network structure-based penalized function to calculate the global Lagrange multiplier.
Theoretically, we demonstrate the Lagrange multipliers defined by the penalized function are uniformly equivalent to the global Lagrange multiplier with probability tending to one, and hence prove that the distributed empirical log-likelihood ratio statistic is asymptotically standard chi-squared.
In Section \ref{sec:algorithm}, we propose two specific algorithms PCM and MAOM for the purpose of solving the optimization problem with the fused penalty in the decentralized distributed manner.
Meanwhile, we establish the convergence property for the proposed algorithms, and further prove the linear convergence of MAOM in some specific network structures.
In Section \ref{sec:simulation}, we conduct some simulation studies to demonstrate the performance of the proposed algorithms.
Analyses of real data for census income and Ford gobike datasets are provided in Section \ref{sec:real data}.

\section{Methodology and Asymptotic Properties}\label{sec:framework}

Let $\bm X_{i,j}$, $\coti$, $\cotj$, be $d$-dimensional independent and identically distributed observations and $\bm\theta = (\theta_1,\ldots,\theta_p)^\top$ a $p$-dimensional parameter with support $\bm\Theta$.
For an $r$-dimensional estimating function $\bm g(\bm X;\bm\theta) = \leftsecond g_1(\bm X;\bm\theta),\ldots,g_r(\bm X;\bm\theta) \rightsecond^\top$, the information for the model parameter $\bm\theta$ is collected by the unbiased moment condition
\begin{equation*}
\oE \leftsecond \bm g(\bm X_{i,j};\bm\theta_0) \rightsecond = \bm 0,~\coti,~\cotj,
\end{equation*}
where $\bm\theta_0\in\bm\Theta$ is the unknown true value.
From \cite{qin1994empirical}, for a given $\bm\theta$, if $\bm 0$ is inside the the convex hull of the points $\bm g(\bm X_{1,1};\bm\theta),\ldots,\bm g(\bm X_{K,n};\bm\theta)$, the empirical log-likelihood ratio statistic is
\begin{equation*}
\max_{\bm\lambda} 2\sum_{i=1}^K \sum_{j=1}^n \log \leftsecond 1+ \bm\lambda^\top \bm g(\bm X_{i,j};\bm\theta) \rightsecond
\end{equation*}
It is direct to calculate the Lagrange multiplier if the observations $\bm X_{i,j}$, $\coti$, $\cotj$, are stored in a single machine.
When the sample size, $N=Kn$, is extraordinarily large, it is impractical to store the whole dataset and compute the Lagrange multiplier in a single machine.
Moreover, the distributed statistical approaches with a central machine also suffers from serious limitations.
(a) Privacy protection: each worker machine sends local statistic to the central machine.
If the central machine is conquered, the attacker has chance to acquire private data information from every worker machines;
(b) Network stability: the centralized network heavily relies on the central machine.
If the central machine stops working, the distributed algorithm cannot continue to work.
(c) Bandwidth requirement: all the information exchanges carry out between the central machine and worker machines.
The incoming and outgoing network bandwidth of the central machine need to be sufficient wide, especially when the number of worker machines diverges.
In this paper, we advocate the idea of decentralized distributed computation.
In this manner, there is no central machine involved for computation, and all the computation-related communications should occur only between directly connected nodes.

The difficulty of developing empirical likelihood in a decentralized distributed manner concentrates on the computation of the Lagrange multiplier.
The Lagrange multiplier calculated with the whole dataset is
\begin{equation}\label{def:lambda*}
\widehat{\bm\lambda}^* = \arg\min_{\bm\lambda} \ell(\bm\lambda;\bm\theta),
\end{equation}
where
\begin{equation*}
\ell(\bm\lambda;\bm\theta) = \sumi \ell_i(\bm\lambda;\bm\theta) = - 2 \sumi\sumj \log \leftsecond 1 + \bm\lambda^\top \bm g(\bm X_{i,j},\bm\theta) \rightsecond.
\end{equation*}
Hereinafter, we denote $\ell(\bm\lambda;\bm\theta)$ and $\ell_i(\bm\lambda;\bm\theta)$ as $\ell(\bm\lambda)$ and $\ell_i(\bm\lambda)$, $\coti$, respectively, whenever no confusion arises.

When calculating the Lagrange multiplier in the decentralized distributed manner, each node performs calculation according to the exclusive data and their neighbors’ information.
If the neighbours' information is not utilized, the $i$-th node calculates the local Lagrange multiplier $\widehat{\bm\lambda}_i^* = \arg\min_{\bm\lambda} \ell_i(\bm\lambda)$ for $\coti$, and the global statistics is constructed as $\sum_{i=1}^K - \ell_i(\widehat{\bm\lambda}_i^*)$.
However, the asymptotic distribution of $\sum_{i=1}^K - \ell_i(\widehat{\bm\lambda}_i^*)$ is not standard chi-squared as $K$ diverges, and the power of the inference by $\sum_{i=1}^K - \ell_i(\widehat{\bm\lambda}_i^*)$ is much lower comparing with that by $-\ell(\widehat{\bm\lambda}^*)$ even for a finite $K$.
The reason may be that $\widehat{\bm\lambda}_i^*$, $\coti$, are calculated separately without making use of data information from other nodes, and this lead to the change of the asymptotic distribution.
To address the problem, and make full use of the neighbours' information, we propose a constrained optimization problem
\begin{equation*}
\begin{aligned}
&\text{minimize} \quad \sumi \ell_i(\bm\lambda_i)\\
&\text{subject to} \quad \bm\lambda_i = \bm\lambda_{i'}, ~\text{for}~ i=1,\ldots,K,~\text{and}~ i'\in\mathcal{N}_i,
\end{aligned}
\end{equation*}
where $\mathcal{N}_i$ represents the set of the neighboring nodes of the $i$-th node, which is pre-specified by the communication network.
To solve the constrained optimization problem, we define a penalized function as follows
\begin{equation}\label{eq:penalty}
\ell_{penalty}(\bm\Lambda) = \sumi \ell_i(\bm\lambda_i) + \sumi \sum_{\stackrel{i'\in \mathcal{N}_i}{i'>i}} P_{\eta_n} (\bm\lambda_i - \bm\lambda_{i'}),
\end{equation}
where $\bm\Lambda = (\bm\lambda_1^\top,\ldots,\bm\lambda_K^\top)^\top \in \mathbb{R}^{Kr}$,
and $P_{\eta_n}(\cdot)$ is a penalty function with the tuning parameter $\eta_n > 0$.
Note that the objective function in \eqref{eq:penalty} consists of two parts: the first term implies that the calculation of $\bm\lambda_i$ uses the exclusive data stored in the $i$-th node for $i=1,\ldots,K$; the second term makes that the calculation of $\bm\lambda_i$ also takes advantage of the information from the neighbours of the $i$-th node.
The penalty term shrinks $\bm\lambda_i - \bm\lambda_{i'}$ to zero vectors for $\coti$, and $i'\in \mathcal{N}_i$.
In other words, with a suitably chosen $\eta_n$, the penalty term facilitates the equivalence between $\bm\lambda_i$ and $\bm\lambda_{i'}$ if the $i$-th node and the $i'$-th node are directly connected in the network.
Therefore, in a connected network (any two of the nodes are directly connected by an edge or indirectly connected by a path), the penalty term guarantees that all the $\bm\lambda_i$ are equal to the minimizer of $\ell_{penalty}(\bm\Lambda)$.
For the choice of $P_{\eta_n} (\cdot)$, many penalty functions are available, such as LASSO \citep{tibshirani1996lasso}, SCAD \citep{fan2001scad}, and MCP \citep{zhang2010mcp}.
In this paper, we adopt the lasso penalty because it not only has a simpler form, but also tends to over-shrink large coefficients \citep{ma2017concave}.
The tendency to over-shrink large coefficients is the major drawback in heterogeneity analyses.
But it is beneficial to the calculation in this paper because the main purpose of adding the penalty term is shrinking all the differences between $\bm\lambda_i$ and $\bm\lambda_{i'}$ to zero vectors.
Such a penalty term is always mentioned as fused lasso \citep{tibshirani2005sparsity,tang2016fused}.
We specialize the objective function
\begin{equation}\label{eq:lasso}
\ell_{lasso}(\bm\Lambda) = \sumi \ell_i(\bm\lambda_i) + \sumi \sum_{\stackrel{i'\in \mathcal{N}_i}{i'>i}} \eta_n \| \bm\lambda_i - \bm\lambda_{i'}\|_2.
\end{equation}
The corresponding solution is defined as
\begin{equation}\label{opt:lasso}
\widehat{\bm\Lambda} = (\widehat{\bm\lambda}_1^\top,\ldots, \widehat{\bm\lambda}_K^\top)^\top = \arg\min_{\bm\Lambda} \ell_{lasso}(\bm\Lambda).
\end{equation}
To establish our asymptotic results, we need the following assumptions.

\begin{assumption}\label{assum:connected}
The network is connected, that is, there exists a path between any two nodes $i$ and $i'$ for $i\neq i'$ and $i,i' = 1,\ldots,K$.
\end{assumption}

\begin{assumption}\label{assum:posi-def}
$\oE \leftsecond \bm g(\bm X;\bm\theta_0) \bm g(\bm X;\bm\theta_0)^\top \rightsecond$ is finite and positive definite.
\end{assumption}

Assumption \ref{assum:connected} makes sure that there is no offline node in the network, and Assumption \ref{assum:posi-def} is mild.
Let $\widehat{\bm\Lambda}^* = (\widehat{\bm\lambda}^{*\top},\ldots, \widehat{\bm\lambda}^{*\top})^\top \in \mathbb{R}^{Kr}$ with $\widehat{\bm\lambda}^*$ being defined in \eqref{def:lambda*}.
We have the following theorems.
\begin{theorem}\label{thm:equal}
Suppose Assumption \ref{assum:connected} and \ref{assum:posi-def} hold.
For $\bm\theta \in \mathbb{R}^p$, under the condition $\eta_n/N^2 \to\infty$, we have
\begin{equation}\label{eq:equal}
\Pr (\widehat{\bm\Lambda} = \widehat{\bm\Lambda}^*) \to 1,~\text{as}~n\to\infty.
\end{equation}
The condition of $\eta_n$ can be relaxed to $\eta_n/K\sqrt{n\log K} \to\infty$ if the parameter $\bm\theta$ satisfies $\|\bm\theta - \bm\theta_0\|_2 = O(N^{-1/2})$.
\end{theorem}

\begin{theorem}\label{thm:convergence}
Under assumptions and conditions of Theorem \ref{thm:equal}, we have
\begin{equation*}
- \sumi \ell_i(\widehat{\bm\lambda}_i;\bm\theta_0) \overset{d}{\to} \chi_{(r)}^2,~\text{as}~n\to\infty,
\end{equation*}
where ``$\overset{d}{\to}$'' denotes the convergence in distribution.
\end{theorem}

Theorem \ref{thm:equal} guarantees that $\widehat{\bm\lambda}_i$, $\coti$, defined by \eqref{opt:lasso} are uniformly equivalent to $\widehat{\bm\lambda}^*$ defined by \eqref{def:lambda*} with probability tending to $1$.
This implies that the distributed log-likelihood ratio statistic $- \sumi \ell_i(\widehat{\bm\lambda}_i;\bm\theta_0)$ is asymptotically standard chi-squared.
However, it is not an easy task to calculate \eqref{opt:lasso} over a decentralized network.
For example, \cite{tang2016fused} uses linear transformation method to calculate estimators in linear model with fused lasso penalty.
The transformation contributes to calculating the estimators with data stored in a single machine.
However, it introduces coupling structure to the objective function, which cannot be solved in the distributed manner.
This motivates us to develop some specific algorithms.

\section{Algorithms}\label{sec:algorithm}

Before designing decentralized distributed algorithms for solving \eqref{opt:lasso}, some useful notations for the network structure should be introduced.
We consider the communication network which has an undirected connected graph structure with $K$ nodes and $M$ edges, $\mathcal{G} = \{\mathcal{V},\mathcal{E}\}$, where $\mathcal{V}:=\{1,\ldots,K\}$ denotes the set of nodes and $\mathcal{E}:=\{e_{i,i'}:\, \text{node } i \text{ and node } i' \text{ are connected by an edge } e_{i,i'} \text{ with } i<i'\}$ denotes the set of edges.
We take the (oriented) incidence matrix $\mathbf{A} \in \mathbb{R}^{M\times K}$ to represent the structure of $\mathcal{G}$ following the definition from \cite{chow2016expander} and \cite{wei2012distributed}.
Each row of matrix $\mathbf{A}$ corresponds to an edge in the graph and each column of matrix $\mathbf{A}$ represents a node.
If the $l$-th row of $\mathbf{A}$ is corresponding to the edge $e_{i,i'}$, we denote $\mathbf{A}_{l \bullet}$ as the $l$-th row of $\mathbf{A}$ satisfying $\mathbf{A}_{l \bullet} = (\bm e_i - \bm e_{i'})^\top$, where $\bm e_i \in \mathbb{R}^K$ is the vector whose $i$-th element is $1$ and the remaining ones are $0$.
For instance, for $\mathcal{G}_0 = \{ \mathcal{V}_0,\mathcal{E}_0 \}$ with $\mathcal{V}_0 = \{1,2,3,4\}$ and $\mathcal{E}_0 = \{e_{1,2},e_{1,3},e_{2,3},e_{2,4},e_{3,4}\}$, the corresponding incidence matrix $\mathbf{A}_0$ is defined as
\begin{equation*}
\mathbf{A}_0 = \left(
\begin{aligned}
1 && -1 && 0 && 0\\
1 && 0 && -1 && 0\\
0 && 1 && -1 && 0\\
0 && 1 && 0 && -1\\
0 && 0 && 1 && -1
\end{aligned}
\right).
\end{equation*}
In the connected graph $\mathcal{G} = \{\mathcal{V},\mathcal{E}\}$ with incidence matrix $\mathbf{A}$, we propose two algorithms to calculate the Lagrange multiplier of the distributed empirical likelihood in the decentralized distributed manner.

It is difficult to compute $\widehat{\bm\Lambda}$ directly from \eqref{opt:lasso} in a decentralized distributed manner because of the non-separable penalty term.
The penalty term encourages information exchange among all the nodes.
However, the communications can occur only between directly connected nodes limited by the network structure.
It is a beneficial approach to reparameterize $\bm\lambda_i$ for $\coti$, and separate $\bm\lambda_i - \bm\lambda_{i'}$ for $\coti$, $i'\in\mathcal{N}_i$ for designing the distributed algorithm.

\subsection{The Pairwise Copy Method}

Motivated by \cite{hallac2015network}, we regard edges $e_{i,i'}\in\mathcal{E}$ as independent constraints, and introduce copies of $\bm\lambda_i$ and $\bm\lambda_{i'}$ as $\bm c_{i,i'}$ and $\bm c_{i',i}$, respectively.
By introducing intermediate parameters, the non-separable structure can be effectively reparameterized, which weakens the directly constraints between $\bm\lambda_i$ and $\bm\lambda_{i'}$.
This reparameterization technique is also adopted by \cite{shi2014linear}.
Based on the copies, we propose the Pairwise Copy Method (PCM).
The minimization of \eqref{eq:lasso} is equivalent to the constrained optimization problem
\begin{equation}\label{eq:opt1}
\begin{aligned}
&\text{minimize} \quad \sumi \ell_i(\bm\lambda_i) + \sum_{e_{i,i'}\in \mathcal{E}} \eta_n \| \bm c_{i,i'} - \bm c_{i',i}\|_2\\
&\text{subject to} \quad \bm\lambda_i = \bm c_{i,i'},\, \bm\lambda_{i'} = \bm c_{i',i},~\text{for}~(i,i') ~\text{with}~ e_{i,i'}\in \mathcal{E}.
\end{aligned}
\end{equation}
In the following, we abbreviate the summation $\sum_{(i,i') ~\text{with}~ e_{i,i'}\in\mathcal{E}}$ as $\sum_{e_{i,i'}\in \mathcal{E}}$ for simplicity of notation.
By the augmented Lagrangian method, given a tuning parameter $\rho>0$, the solution of \eqref{eq:opt1} can be obtained by minimizing
\begin{equation}\label{eq:AL1}
\begin{aligned}
&\mathcal{L}_{PCM} (\bm\Lambda,\bm C,\bm V)\\
=& \sumi \ell_i(\bm\lambda_i) + \sum_{e_{i,i'}\in \mathcal{E}} \eta_n \| \bm c_{i,i'} - \bm c_{i',i}\|_2 + \sum_{e_{i,i'}\in \mathcal{E}} \leftsecond \bm v_{i,i'}^\top (\bm\lambda_i - \bm c_{i,i'}) + \bm v_{i',i}^\top (\bm\lambda_{i'} - \bm c_{i',i}) \rightsecond\\
&+ \frac{\rho}{2} \sum_{e_{i,i'}\in \mathcal{E}} \leftsecond \| \bm\lambda_i - \bm c_{i,i'} \|_2^2 + \| \bm\lambda_{i'} - \bm c_{i',i} \|_2^2 \rightsecond
\end{aligned}
\end{equation}
in terms of $\bm\Lambda$, $\bm C$ and $\bm V$, where $\bm v_{i,i'}$ and $\bm v_{i',i}$ are dual vectors for $(i,i')$ with $e_{i,i'} \in \mathcal{E}$, and $\bm C$, $\bm V$ are the corresponding $2Mr$-dimensional vectors composed of $\bm c_{i,i'}$, $\bm c_{i',i}$ and $\bm v_{i,i'}$, $\bm v_{i',i}$, respectively, for $(i,i')$ with $e_{i,i'} \in\mathcal{E}$.
We now establish the computational algorithm based on the
ADMM by minimizing $\mathcal{L}_{PCM} (\bm\Lambda,\bm C,\bm V)$ defined in \eqref{eq:AL1}.
It consists of steps for iteratively updating $\bm C$, $\bm\Lambda$ and $\bm V$.
Given $(\bm\Lambda^\iterO, \bm C^\iterO, \bm V^\iterO)$, the update in the $t$-th iteration, we derive the update strategy for $(\bm\Lambda^\iterN, \bm C^\iterN, \bm V^\iterN)$.

First, we update $\bm C^{(t+1)}$ by calculating $\bm C^{(t+1)} = \arg\min_{\bm C} \mathcal{L}_{PCM} (\bm\Lambda^{(t)}, \bm C, \bm V^{(t)})$.
For $(i,i')$ with $e_{i,i'}\in \mathcal{E}$, we have
\begin{equation*}
\begin{aligned}
\left( \bm c_{i,i'}^{(t+1)},\bm c_{i',i}^{(t+1)} \right) = \arg\min_{\left( \bm c_{i,i'},\bm c_{i',i} \right)} {}&{} \big\{ \eta_n \| \bm c_{i,i'} - \bm c_{i',i}\|_2 + \bm v_{i,i'}^{(t)\top} (\bm\lambda_i^{(t)} - \bm c_{i,i'}) + \bm v_{i',i}^{(t)\top} (\bm\lambda_{i'}^{(t)} - \bm c_{i',i})\\
{}&{}+ \frac{\rho}{2} \| \bm\lambda_i^{(t)} - \bm c_{i,i'} \|_2^2 + \frac{\rho}{2} \| \bm\lambda_{i'}^{(t)} - \bm c_{i',i} \|_2^2 \big\}.
\end{aligned}
\end{equation*}

Second, we update $\bm\Lambda^{(t+1)}$ by solving
$$
\partial \mathcal{L}_{PCM} (\bm\Lambda,\bm C^{(t+1)},\bm V^{(t)}) / \partial \bm\Lambda = \bm 0.
$$
Hence, for $i$-th node, $\coti$, $\bm\lambda_i^{(t+1)}$ satisfies
\begin{equation}\label{PCM:Lambda}
\nabla \ell_i(\bm\lambda_i^{(t+1)}) + \sum_{i'\in \mathcal{N}_i} \bm v_{i,i'}^{(t)} + \rho \sum_{i'\in \mathcal{N}_i} ( \bm\lambda_i^{(t+1)} - \bm c_{i,i'}^{(t+1)}) = \bm 0.
\end{equation}

Finally, the dual vector $\bm V^{(t+1)}$ is easy to compute from the general theory of ADMM.
For $(i,i')$ with $e_{i,i'}\in \mathcal{E}$, we have
\begin{equation}\label{PCM:V}
\bm v_{i,i'}^{(t+1)} = \bm v_{i,i'}^{(t)} + \rho (\bm\lambda_i^{(t+1)} - \bm c_{i,i'}^{(t+1)}),\quad \bm v_{i',i}^{(t+1)} = \bm v_{i',i}^{(t)} + \rho (\bm\lambda_{i'}^{(t+1)} - \bm c_{i',i}^{(t+1)}).
\end{equation}

When updating $\bm C^\iterN$, the optimization problem has a closed-form solution, which is derived in the following lemma.
This is an important factor when we consider the update order of $\bm C$ and $\bm\Lambda$.
The update procedure for $\bm C$ has a closed-form solution, while that for $\bm\Lambda$ demands numerical computation which may be less precise.
Firstly updating $\bm\Lambda$ in each round of iteration leads to a slower convergence rate because the later calculations are based on the previous results.
For these reasons, we update $\bm C$ antecedent to $\bm\Lambda$ in each round of iteration.

\begin{lemma}\label{lemma:closed-form PCM}
Given $\bm\Lambda^\iterO$ and $\bm V^\iterO$, the update procedure for $\bm C^\iterN$ has a closed-form solution in PCM, that is, for $(i,i')$ with $e_{i,i'}\in \mathcal{E}$
\begin{equation}\label{PCM:C}
\begin{aligned}
\bm c_{i,i'}^\iterN =& \omega^\iterO (\bm\lambda_i^\iterO + \rho^{-1} \bm v_{i,i'}^\iterO) + (1-\omega^\iterO) (\bm\lambda_{i'}^\iterO + \rho^{-1} \bm v_{i',i}^\iterO),\\
\bm c_{i',i}^\iterN =& (1-\omega^\iterO) (\bm\lambda_i^\iterO + \rho^{-1} \bm v_{i,i'}^\iterO) + \omega^\iterO (\bm\lambda_{i'}^\iterO + \rho^{-1} \bm v_{i',i}^\iterO),
\end{aligned}
\end{equation}
where
\begin{equation*}
\omega^\iterO = \max \leftsecond 1- \frac{\eta_n}{\| \rho \bm\lambda_i^\iterO - \rho \bm\lambda_{i'}^\iterO + \bm v_{i,i'}^\iterO - \bm v_{i',i}^\iterO \|_2}, 0.5 \rightsecond.
\end{equation*}
\end{lemma}

From conditions of Theorem \ref{thm:equal}, we notice that $\eta_n$ diverges with $n$ in a relatively fast rate.
Hence, in most cases, the value of $\omega^\iterO$ defined in Lemma \ref{lemma:closed-form PCM} is $0.5$.
This implies the equivalence between $\bm c_{i,i'}$ and $\bm c_{i',i}$, and further facilitates the equivalence between $\bm\lambda_i$ and $\bm\lambda_{i'}$.

As alluded to above, it can be shown that the update procedure for $\bm\lambda_i$ only requires exclusive data stored in the $i$-th node and information of $\{ \bm v_{i,i'},\bm c_{i,i'}:\, i'\in \mathcal{N}_i \}$, and the update procedures for $\bm v_{i,i'}$ and $\bm c_{i,i'}$ only require information of $\{ \bm\lambda_i, \bm\lambda_{i'}, \bm v_{i,i'}, \bm c_{i,i'} \}$.
Hence, the algorithm can be carried out in the decentralized distributed manner.
For the first round of iteration, the initial values $\bm\Lambda^{(0)}$, $\bm C^{(0)}$, and $\bm V^{(0)}$ can be set as zero vectors.
We track the progress of PCM based on the primal residual $\bm r_1$ and dual residual $\bm s_1$, and stop the algorithm when the residuals are close to zero.
The specific expression for the primal and dual residuals are $\bm r_1^\iterO = \widetilde{\mathbf{A}}_{LR} \bm\Lambda^\iterO - \bm C^\iterO$ and $\bm s_1^\iterO = \rho \widetilde{\mathbf{A}}_{LR}^\top (\bm V^{(t-1)} - \bm V^\iterO)$, respectively, where the explicit expression for $\widetilde{\mathbf{A}}_{LR}$ is defined in the proof
section.
Based on the explanation above, we establish the algorithm for PCM.

\begin{algorithm}[H]
\caption{The pairwise copy method (PCM)}\label{alg:PCM}
\begin{algorithmic}[1]
\Require The initial values $\bm\Lambda^{(0)}$, $\bm C^{(0)}$, and $\bm V^{(0)}$, the tolerances for the primal and
dual residuals $\epsilon^{pri}$ and $\epsilon^{dual}$.

\State Set $t=0$, and initial values for the primal and
dual residuals $\| \bm r_1^{(0)} \|_2 =\infty$ and $\| \bm s_1^{(0)} \|_2 =\infty$;

\While {convergence criterion is not met, that is, $\| \bm r_1^\iterO \|_2 > \epsilon^{pri}$ or $\| \bm s_1^\iterO \|_2 > \epsilon^{dual}$}

\State According to \eqref{PCM:C}, for $(i,i')$ with $e_{i,i'}\in \mathcal{E}$, calculate
\begin{equation*}
\begin{aligned}
\bm c_{i,i'}^\iterN =& \omega^\iterO (\bm\lambda_i^\iterO + \rho^{-1} \bm v_{i,i'}^\iterO) + (1-\omega^\iterO) (\bm\lambda_{i'}^\iterO + \rho^{-1} \bm v_{i',i}^\iterO),\\
\bm c_{i',i}^\iterN =& (1-\omega^\iterO) (\bm\lambda_i^\iterO + \rho^{-1} \bm v_{i,i'}^\iterO) + \omega^\iterO (\bm\lambda_{i'}^\iterO + \rho^{-1} \bm v_{i',i}^\iterO),
\end{aligned}
\end{equation*}
where $\omega^\iterO$ is defined in Lemma \ref{lemma:closed-form PCM};

\State According to \eqref{PCM:Lambda}, for $\coti$, calculate $\bm\lambda_i^\iterN$ by
\begin{equation*}
\nabla \ell_i(\bm\lambda_i^\iterN) + \sum_{i'\in \mathcal{N}_i} \bm v_{i,i'}^\iterO + \rho \sum_{i'\in \mathcal{N}_i} ( \bm\lambda_i^\iterN - \bm c_{i,i'}^\iterN) = \bm 0;
\end{equation*}

\State According to \eqref{PCM:V}, for $(i,i')$ with $e_{i,i'}\in \mathcal{E}$, calculate
\begin{equation*}
\bm v_{i,i'}^{(t+1)} = \bm v_{i,i'}^{(t)} + \rho (\bm\lambda_i^{(t+1)} - \bm c_{i,i'}^{(t+1)}),\quad \bm v_{i',i}^{(t+1)} = \bm v_{i',i}^{(t)} + \rho (\bm\lambda_{i'}^{(t+1)} - \bm c_{i',i}^{(t+1)});
\end{equation*}

\State Compute the primal and dual residuals $\bm r_1^\iterN$ and $\bm s_1^\iterN$;

\State Set $t = t+1$;

\EndWhile

\Ensure $(\bm\Lambda^\iterO, \bm C^\iterO, \bm V^\iterO)$.

\end{algorithmic}
\end{algorithm}

Denote the KKT point for \eqref{eq:opt1} as $\bm U_{1*} = (\bm\Lambda_{1*}^\top, \bm C_*^\top, \bm V_*^\top)^\top$.
Algorithm \ref{alg:PCM} ensures that $\bm U_1^{(t)} = (\bm\Lambda^{(t)\top}, \bm C^{(t)\top}, \bm V^{(t)\top})^\top$ converges to $\bm U_{1*}$ in a few tens of iterations, which is derived in the following theorem.

\begin{theorem}\label{thm:KKTpoint1}
Under assumptions and conditions of Theorem \ref{thm:equal}, we have
\begin{equation*}
\lim_{t\to\infty} \| \bm U_1^{(t)} - \bm U_{1*} \|_2 = 0.
\end{equation*}
\end{theorem}

Theorem \ref{thm:convergence} ensures that the distributed log-likelihood ratio statistic $- \sumi \ell_i(\widehat{\bm\lambda}_i;\bm\theta_0)$ has the same asymptotic distribution and hence it is asymptotically standard chi-squared.
Theorem \ref{thm:KKTpoint1} guarantees that $\bm\lambda_i^{(t)}$, $\coti$, calculated by Algorithm \ref{alg:PCM} converge to $\widehat{\bm\lambda}^*$ uniformly as $t\to\infty$.
Therefore, we have the following corollary.

\begin{corollary}\label{coro:PCM}
Under assumptions and conditions of Theorem \ref{thm:equal}, we have
\begin{equation*}
\lim_{t\to\infty} -\sumi \ell_i(\bm\lambda_i^{(t)};\bm\theta_0) \overset{d}{\to} \chi_{(r)}^2,~\text{as}~n\to\infty,
\end{equation*}
where $\bm\lambda_i^{(t)}$ is calculated according to Algorithm \ref{alg:PCM}.
\end{corollary}

It is worth noting that there is only a single algorithm parameter $\rho$ in Algorithm \ref{alg:PCM}, and the algorithm is shown to converge for all $\rho>0$.
However, $\rho$ has a direct impact on the convergence rate of the algorithm, and inadequate tuning of this parameter can render the method slow.
By analyzing the magnitude of $\bm U_{1*}$, we recommend the choice of $\rho = n$ to accelerate the convergence speed of PCM.




\subsection{The Modified Approximation Objective Method}

The pairwise copies used in PCM contribute to decoupling of the penalty term, which work well in the decentralized distributed manner.
However, introducing superabundant parameters not only slows down the convergence speed, but also increases the computation cost.
Moreover, even though we improve the convergence performance of PCM by updating $\bm C$ first, the computation burden in each round of iteration is still heavy because the update strategy for $\bm\Lambda$ relies on numerical computation, especially when the dimension of the estimating function grows.
To fix it, we propose the Modified Approximation Objective Method (MAOM), which introduces less parameters and simplifies computation process in each iteration.
In PCM, we introduce $\bm c_{i,i'}$ and $\bm c_{i',i}$ to represent the two values $\bm\lambda_i$ and $\bm\lambda_{i'}$ for each edge $e_{i,i'}$.
This results in $2M$ parameters and $2M$ dual variables in \eqref{eq:AL1}.
Naturally, we can halve the number of parameters by fix only one parameter to one edge.
To be specific, we denote $\bm z_{ii'} = \bm\lambda_i - \bm\lambda_{i'}$ for $(i,i')$ with $e_{i,i'} \in \mathcal{E}$.
The minimization of \eqref{eq:lasso} is equivalent to the constrained optimization problem
\begin{equation}\label{eq:opt2}
\begin{aligned}
& \text{minimize} \quad \sumi \ell_i(\bm\lambda_i) + \sum_{e_{i,i'}\in \mathcal{E}} \eta_n \| \bm z_{ii'} \|_2\\
& \text{subject to} \quad \bm\lambda_i - \bm\lambda_{i'} = \bm z_{ii'},~\text{for}~ (i,i') ~\text{with}~ e_{i,i'}\in \mathcal{E}.
\end{aligned}
\end{equation}
By the augmented Lagrangian method, given a tuning parameter $\rho>0$, the solution of \eqref{eq:opt2} can be obtained by minimizing
\begin{equation}\label{eq:AL2}
\begin{aligned}
\mathcal{L}_{MAOM} (\bm\Lambda, \bm Z, \bm T) = & \sumi \ell_i(\bm\lambda_i) + \sum_{e_{i,i'}\in \mathcal{E}} \eta_n \| \bm z_{ii'} \|_2 + \sum_{e_{i,i'}\in \mathcal{E}} \bm t_{ii'}^\top (\bm\lambda_i - \bm\lambda_{i'} - \bm z_{ii'})\\
& + \frac{\rho}{2} \sum_{e_{i,i'}\in \mathcal{E}} \|\bm\lambda_i - \bm\lambda_{i'} - \bm z_{ii'}\|_2^2,
\end{aligned}
\end{equation}
where $\bm t_{ii'}$ is the dual vector for $(i,i')$ with $e_{i,i'}\in \mathcal{E}$.

In comparison to PCM, MAOM reduces the number of parameters by half.
However, as what is illustrated in PCM, the extra parameters decouple the penalty term and ensure the feasibility of ADMM in the decentralized network.
It results in the non-separable structure in the problem to reduce the number of parameters.
Hence, the optimization problem cannot be directly solved in a decentralized distributed manner.
Moreover, the computation of $\bm\Lambda$ in each round of iteration is still time-consuming due to the structure of $\sum_{i=1}^K \ell_i(\bm\lambda_i)$.
To address the problems mentioned above, in each round of iteration, we replace the objective function by its second-order approximation, and modify it by adding a strict convex quadratic regularization function.
After meticulously designing the regularization function, the optimization problem can be solved in the decentralized distributed manner, and has the closed-form solution in each round of iteration.
We adjust the modified approximation objective function according to the current value of $\bm\Lambda$ in each round of iteration since fixed approximation leads to a wrong target.
It is worth noting that the modified approximation does not decrease the convergence rate of the algorithm.
Moreover, the convergence property for the varying objective function, which depends on the current update values, has not been previously proven within the framework of ADMM.
The details for the modified approximation and the corresponding update strategy are illustrated below.

For simplicity of expression, we redefine \eqref{eq:AL2} in the matrix form.
Let $\bm Z = \{\bm z_{ii'}:\, e_{i,i'} \in\mathcal{E} \}^\top$ and $\bm T = \{\bm t_{ii'}:\, e_{i,i'} \in\mathcal{E} \}^\top$.
\eqref{eq:AL2} can be represented as
\begin{equation}\label{eq:AL3}
\mathcal{L}_{MAOM} (\bm\Lambda, \bm Z, \bm T) = \mathcal{L}(\bm\Lambda) + \sum_{e_{i,i'}\in \mathcal{E}} \eta_n \| \bm z_{ii'} \|_2 + \bm T^\top (\widetilde{\mathbf{A}} \bm\Lambda - \bm Z) + \frac{\rho}{2} \|\widetilde{\mathbf{A}} \bm\Lambda - \bm Z\|_2^2,
\end{equation}
where $\mathcal{L}(\bm\Lambda) = \sumi \ell_i(\bm\lambda_i)$, $\widetilde{\mathbf{A}} = \mathbf{A} \otimes \mathbf{I}_r$, $\otimes$ denotes the Kronecker product, and $\mathbf{I}_r \in \mathbb{R}^{r\times r}$ is the identity matrix.
Similarly, we minimize $\mathcal{L}_{MAOM} (\bm\Lambda, \bm Z, \bm T)$ defined in \eqref{eq:AL3} by alternatively updating $\bm Z$, $\bm\Lambda$ and $\bm T$.
Before the formal illustration of the algorithm, we introduce the adaptive modified approximation.
Given $\bm\Lambda^{(t)}$, we replace the function $\mathcal{L}(\bm\Lambda)$ by
\begin{equation}\label{eq:MAL}
\begin{aligned}
\widehat{\mathcal{L}}^{(t)} (\bm\Lambda;\widetilde{\mathbf{Q}}) ={}&{} \mathcal{L} (\bm\Lambda^{(t)}) + (\bm\Lambda - \bm\Lambda^{(t)})^\top \nabla \mathcal{L} (\bm\Lambda^{(t)})\\
{}&{} + \frac{1}{2} (\bm\Lambda - \bm\Lambda^{(t)})^\top \nabla^2 \mathcal{L} (\bm\Lambda^{(t)}) (\bm\Lambda - \bm\Lambda^{(t)}) + \frac{1}{2} (\bm\Lambda - \bm\Lambda^{(t)})^\top \widetilde{\mathbf{Q}} (\bm\Lambda - \bm\Lambda^{(t)}),
\end{aligned}
\end{equation}
where $\widetilde{\mathbf{Q}} \succ \mathbf{0}$ is a positive definite symmetric matrix, which is specified later.
The last term at the right hand side of \eqref{eq:MAL} is a strict convex quadratic regularization function, which can be regard as the proximal function in the sense of $\widetilde{\mathbf{Q}}$-norm.
It enforces that the iterations for $\bm\Lambda^\iterN$ do not oscillate wildly.
If we set $\widetilde{\mathbf{Q}} = \tau \mathbf{I}_{Kr}$ with $\tau>0$, the proximal function degenerates to the traditional proximal function introduced by \cite{rockafellar1976monotone}.
Additionally, the specially designed $\widetilde{\mathbf{Q}}$ contributes to solving the problem in the decentralized distributed manner.
Generally speaking, given $\bm\Lambda^{(t)}$, we should construct the modified approximation objective function and calculate the optimal solution several times to get the precise update $\bm\Lambda^{(t+1)}$.
However, it is worth noting that the solution of the second-order approximation function converges to the optimal solution quadratically according to the convergence property of Newton-Raphson algorithm.
Meanwhile, the proximal point algorithm converges linearly, and the fastest convergence rate of ADMM is also linear \citep{deng2016global}.
Intuitively, we can directly replace $\mathcal{L}(\bm\Lambda)$ by $\widehat{\mathcal{L}}^{(t)} (\bm\Lambda)$ in the $(t+1)$-th iteration without slowing down the convergence rate.
In the following, we illustrate the computation procedure, and specify the explicit form of the positive definite symmetric matrix $\widetilde{\mathbf{Q}}$.

Given $(\bm\Lambda^{(t)}, \bm Z^{(t)}, \bm T^{(t)})$, we provide the update strategy for $(\bm\Lambda^{(t+1)}, \bm Z^{(t+1)}, \bm T^{(t+1)})$.
First, we update $\bm Z^{(t+1)}$ by calculating $\bm Z^{(t+1)} = \arg\min_{\bm Z} \mathcal{L}_{MAOM} (\bm\Lambda^{(t)}, \bm Z, \bm T^{(t)})$.
For $(i,i')$ with $e_{i,i'}\in \mathcal{E}$, we have
\begin{equation*}
\begin{aligned}
\bm z_{ii'}^\iterN = {}&{} \arg\min_{\bm z_{ii'}} \leftsecond \eta_n \| \bm z_{ii'} \|_2 + \bm t_{ii'}^{(t)\top} (\bm\lambda_i^{(t)} - \bm\lambda_{i'}^{(t)} - \bm z_{ii'}) + \frac{\rho}{2} \| \bm\lambda_i^{(t)} - \bm\lambda_{i'}^{(t)} - \bm z_{ii'} \|_2^2 \rightsecond\\
= {}&{} \arg\min_{\bm z_{ii'}} \leftsecond \eta_n \| \bm z_{ii'} \|_2 + \frac{\rho}{2} \| \bm\lambda_i^{(t)} - \bm\lambda_{i'}^{(t)} + \rho^{-1} \bm t_{ii'}^{(t)} -\bm z_{ii'} \|_2^2 \rightsecond.
\end{aligned}
\end{equation*}
It has the closed-form solution
\begin{equation}\label{MAOM:Z}
\bm z_{ii'}^\iterN = \operatorname{S} ( \bm\lambda_i^{(t)} - \bm\lambda_{i'}^{(t)} + \rho^{-1} \bm t_{ii'}^{(t)}, \rho^{-1}\eta_n ),
\end{equation}
where $\operatorname{S} ( \bm h, t) = (1- t/\|\bm h \|_2)_{+} \bm h$ is the groupwise thresholding operator with $(x)_{+} = x$ if $x > 0$, and $(x)_{+} = 0$ otherwise.

Second, we update $\bm\Lambda^{(t+1)}$ by calculating $\partial \widehat{\mathcal{L}}_{MAOM}^{(t)} (\bm\Lambda, \bm Z^{(t+1)}, \bm T^{(t)})/ \partial \bm\Lambda = \bm 0$, where
\begin{equation*}
\begin{aligned}
& \widehat{\mathcal{L}}_{MAOM}^{(t)} (\bm\Lambda, \bm Z^{(t+1)}, \bm T^{(t)})\\
=& \widehat{\mathcal{L}}^{(t)} (\bm\Lambda;\widetilde{\mathbf{Q}}) + \sum_{e_{i,i'}\in \mathcal{E}} \eta_n \| \bm z_{ii'}^\iterN \|_2 + \bm T^{(t)\top} (\widetilde{\mathbf{A}} \bm\Lambda - \bm Z^\iterN) + \frac{\rho}{2} \|\widetilde{\mathbf{A}} \bm\Lambda - \bm Z^\iterN\|_2^2,
\end{aligned}
\end{equation*}
and
\begin{equation*}
\begin{aligned}
\frac{\partial}{\partial \bm\Lambda} \widehat{\mathcal{L}}_{MAOM}^{(t)} (\bm\Lambda, \bm Z^{(t+1)}, \bm T^{(t)}) ={}&{} \nabla \mathcal{L} (\bm\Lambda^{(t)}) + \leftsecond \nabla^2 \mathcal{L}( \bm\Lambda^{(t)}) + \widetilde{\mathbf{Q}} \rightsecond (\bm\Lambda - \bm\Lambda^{(t)} )\\
{}&{} + \widetilde{\mathbf{A}}^\top \bm T^{(t)} + \rho \widetilde{\mathbf{A}}^\top ( \widetilde{\mathbf{A}} \bm\Lambda - \bm Z^{(t+1)}).
\end{aligned}
\end{equation*}
According to the adaptive second-order approximation, $\bm\Lambda^{(t+1)}$ has the closed form solution
\begin{equation*}
\begin{aligned}
\bm\Lambda^{(t+1)} = \leftsecond \nabla^2 \mathcal{L}( \bm\Lambda^{(t)}) + \widetilde{\mathbf{Q}} + \rho \widetilde{\mathbf{A}}^\top \widetilde{\mathbf{A}} \rightsecond^{-1} \Big[ {}&{} \leftsecond \nabla^2 \mathcal{L}( \bm\Lambda^{(t)}) + \widetilde{\mathbf{Q}} \rightsecond \bm\Lambda^{(t)} + \rho \widetilde{\mathbf{A}}^\top \bm Z^{(t+1)}\\
{}&{} \leftdot - \nabla \mathcal{L} (\bm\Lambda^{(t)}) - \widetilde{\mathbf{A}}^\top \bm T^{(t)} \rightthird,
\end{aligned}
\end{equation*}
provided $\leftsecond \nabla^2 \mathcal{L}( \bm\Lambda^{(t)}) + \widetilde{\mathbf{Q}} + \rho \widetilde{\mathbf{A}}^\top \widetilde{\mathbf{A}} \rightsecond$ is invertible.

It is worth noting that if the matrix $\leftsecond \nabla^2 \mathcal{L}( \bm\Lambda^{(t)}) + \widetilde{\mathbf{Q}} + \rho \widetilde{\mathbf{A}}^\top \widetilde{\mathbf{A}} \rightsecond$ is not a block diagonal matrix, it is nearly impossible to compute the matrix inverse in a decentralized distributed manner because of the coupled structure \citep{deng2016global,zhang2022learning}.
To address the problem, we design the matrix $\widetilde{\mathbf{Q}}$ to neutralize the effect of the coupled item $\rho \widetilde{\mathbf{A}}^\top \widetilde{\mathbf{A}}$.
Naturally, we can take $\widetilde{\mathbf{Q}} = \widetilde{\mathbf{D}} - \rho \widetilde{\mathbf{A}}^\top \widetilde{\mathbf{A}}$, where $\widetilde{\mathbf{D}} = diag(d_1,\ldots,d_K) \otimes \mathbf{I}_r$ with $d_i>0$ big enough for $\coti$.
The eigenvalues of $\widetilde{\mathbf{A}}^\top \widetilde{\mathbf{A}}$ should be taken into account in order to ensure that $\widetilde{\mathbf{Q}}$ is positive definite.
Let $\mathbf{L} = \mathbf{A}^\top \mathbf{A}$ denote the Laplacian matrix, we have $\widetilde{\mathbf{Q}} = \widetilde{\mathbf{D}} - \rho \mathbf{L} \otimes \mathbf{I}_r$.
According to the definition of $\mathbf{A}$, we have
\begin{equation*}
\mathbf{L}_{ii'} = \left\{
\begin{aligned}
|\mathcal{N}_i|, \quad {}&{} \text{if}~ i=i',\\
-1, \quad {}&{} \text{if}~ e_{i,i'}\in \mathcal{E}~\text{or}~ e_{i',i}\in \mathcal{E} ~ \text{with}~ i\neq i',\\
0, \quad {}&{} \text{if}~ e_{i,i'}\notin \mathcal{E} ~\text{and}~ e_{i',i}\notin \mathcal{E} ~ \text{with}~ i\neq i',
\end{aligned}
\right.
\end{equation*}
where $\mathbf{L}_{ii'}$ denotes the $(i,i')$-th element of $\mathbf{L}$.
Hence, according to the Gerschgorin circle theorem, $\widetilde{\mathbf{Q}}$ is positive definite if $d_i > 2\rho |\mathcal{N}_i|$, for $\coti$.
For simplicity, we take $d_i = 2\rho |\mathcal{N}_i| +1$, $\coti$.

Additionally, $\leftsecond \nabla^2 \mathcal{L}( \bm\Lambda^{(t)}) + \widetilde{\mathbf{D}} \rightsecond$ is invertible because $\nabla^2 \mathcal{L}( \bm\Lambda^{(t)}) \succ \mathbf{0}$ with probability tending to $1$ according to Assumption \ref{assum:posi-def}.
Hence, $\bm\Lambda^{(t+1)}$ has the closed form solution
\begin{equation*}
\begin{aligned}
\bm\Lambda^{(t+1)} = \leftsecond \nabla^2 \mathcal{L}( \bm\Lambda^{(t)}) + \widetilde{\mathbf{D}} \rightsecond^{-1} \Big[ {}&{} \leftsecond \nabla^2 \mathcal{L}( \bm\Lambda^{(t)}) + \widetilde{\mathbf{D}} - \rho \widetilde{\mathbf{A}}^\top \widetilde{\mathbf{A}} \rightsecond \bm\Lambda^{(t)} \\
{}&{} \leftdot + \widetilde{\mathbf{A}}^\top ( \rho \bm Z^{(t+1)} - \bm T^{(t)} ) - \nabla \mathcal{L} (\bm\Lambda^{(t)}) \rightthird.
\end{aligned}
\end{equation*}
Therefore, for $\coti$, we have the update strategy
\begin{equation}\label{MAOM:Lambda}
\begin{aligned}
\bm\lambda_i^{(t+1)} = {}&{} \leftsecond \nabla^2 \ell_i( \bm\lambda_i^{(t)}) + (2 \rho |\mathcal{N}_i| +1) \mathbf{I}_r \rightsecond^{-1} \\
{}&{} \times \bigg[ \leftsecond \nabla^2 \ell_i( \bm\lambda_i^{(t)}) + (\rho |\mathcal{N}_i| +1) \mathbf{I}_r \rightsecond \bm\lambda_i^{(t)} + \rho \sum_{i'\in \mathcal{N}_i} \bm\lambda_{i'}^{(t)}\\
{}&{} + \sum_{e_{i,i'}\in \mathcal{E}} (\rho \bm z_{ii'}^{(t+1)} - \bm t_{ii'}^{(t)}) - \sum_{e_{i'i}\in \mathcal{E}} (\rho \bm z_{i'i}^{(t+1)} - \bm t_{i'i}^{(t)}) - \nabla \ell_i( \bm\lambda_i^{(t)}) \bigg].
\end{aligned}
\end{equation}

Finally, the dual vector $\bm T^{(t+1)}$ can be updated by $\bm T^{(t+1)} = \bm T^{(t)} + \rho (\widetilde{\mathbf{A}} \bm\Lambda^{(t+1)} - \bm Z^{(t+1)})$.
For $(i,i')$ with $e_{i,i'}\in \mathcal{E}$, we have
\begin{equation}\label{MAOM:T}
\bm t_{ii'}^{(t+1)} = \bm t_{ii'}^{(t)} + \rho( \bm\lambda_i^{(t+1)} - \bm\lambda_{i'}^{(t+1)} - \bm z_{ii'}^{(t+1)}).
\end{equation}

As alluded to above, it can be shown that the update procedure for $\bm\lambda_i$ only requires exclusive data stored in the $i$-th node and information of $\{ \bm\lambda_{i'}, \bm z_{ii'},\bm t_{ii'}:\, i'\in \mathcal{N}_i \}$, and the update procedures for $\bm z_{ii'}$ and $\bm t_{ii'}$ only require information of $\{ \bm\lambda_i, \bm\lambda_{i'}, \bm z_{ii'}, \bm t_{ii'} \}$.
Hence, the algorithm can be carried out in the decentralized distributed manner.
For the first round of iteration, the initial values $\bm\Lambda^{(0)}$, $\bm Z^{(0)}$, and $\bm T^{(0)}$ can be set as zero vectors.
We track the progress of PCM based on the primal residual $\bm r_2$ and dual residual $\bm s_2$, and stop the algorithm when the residuals are close to zero.
The specific expression for the primal and dual residuals are $\bm r_2^{(t)} = \widetilde{\mathbf{A}} \bm\Lambda^\iterO - \bm Z^\iterO$ and $\bm s_2^{(t)} = \rho \widetilde{\mathbf{A}}^\top (\bm T^{(t-1)} - \bm T^{(t)})$, respectively.
Based on the explanation above, we establish the algorithm for MAOM.

\begin{algorithm}[H]
\caption{The modified approximation objective method (MAOM)}\label{alg:MAOM}
\begin{algorithmic}[1]
\Require The initial values $\bm\Lambda^{(0)}$, $\bm Z^{(0)}$, and $\bm T^{(0)}$, the tolerances for the primal and
dual residuals $\epsilon^{pri}$ and $\epsilon^{dual}$.

\State Set $t=0$, and initial values for the primal and
dual residuals $\| \bm r_2^{(0)} \|_2 =\infty$ or $\| \bm s_2^{(0)} \|_2 =\infty$;

\While {convergence criterion is not met, that is, $\| \bm r_2^\iterO \|_2 > \epsilon^{pri}$ or $\| \bm s_2^\iterO \|_2 > \epsilon^{dual}$}

\State According to \eqref{MAOM:Z}, for $(i,i')$ with $e_{i,i'}\in \mathcal{E}$, calculate
\begin{equation*}
\bm z_{ii'}^\iterN = \operatorname{S} ( \bm\lambda_i^{(t)} - \bm\lambda_{i'}^{(t)} + \rho^{-1} \bm t_{ii'}^{(t)}, \rho^{-1}\eta_n );
\end{equation*}

\State According to \eqref{MAOM:Lambda}, for $\coti$, calculate
\begin{equation*}
\begin{aligned}
\bm\lambda_i^{(t+1)} = {}&{} \leftsecond \nabla^2 \ell_i( \bm\lambda_i^{(t)}) + (2 \rho |\mathcal{N}_i| +1) \mathbf{I}_r \rightsecond^{-1} \\
{}&{} \times \bigg[ \leftsecond \nabla^2 \ell_i( \bm\lambda_i^{(t)}) + (\rho |\mathcal{N}_i| +1) \mathbf{I}_r \rightsecond \bm\lambda_i^{(t)} + \rho \sum_{i'\in \mathcal{N}_i} \bm\lambda_{i'}^{(t)}\\
{}&{} + \sum_{e_{i,i'}\in \mathcal{E}} (\rho \bm z_{ii'}^{(t+1)} - \bm t_{ii'}^{(t)}) - \sum_{e_{i'i}\in \mathcal{E}} (\rho \bm z_{i'i}^{(t+1)} - \bm t_{i'i}^{(t)}) - \nabla \ell_i( \bm\lambda_i^{(t)}) \bigg];
\end{aligned}
\end{equation*}

\State According to \eqref{MAOM:T}, for $(i,i')$ with $e_{i,i'}\in \mathcal{E}$, calculate
\begin{equation*}
\bm t_{ii'}^{(t+1)} = \bm t_{ii'}^{(t)} + \rho( \bm\lambda_i^{(t+1)} - \bm\lambda_{i'}^{(t+1)} - \bm z_{ii'}^{(t+1)});
\end{equation*}

\State Compute the primal and dual residuals $\bm r_2^\iterN$ and $\bm s_2^\iterN$;

\State Set $t = t+1$;

\EndWhile

\Ensure $(\bm\Lambda^\iterO, \bm Z^\iterO, \bm T^\iterO)$.

\end{algorithmic}
\end{algorithm}

It is worth noting that the update procedure for MAOM has the closed-form solution in each round of iteration, which implies less computation cost.
Denote the KKT point for \eqref{eq:opt2} as $\bm U_{2*} = (\bm\Lambda_{2*}^\top, \bm Z_*^\top, \bm T_*^\top)^\top$.
Algorithm \ref{alg:MAOM} ensures that $\bm U_2^{(t)} = (\bm\Lambda^{(t)\top}, \bm Z^{(t)\top}, \bm T^{(t)\top})^\top$ converges to $\bm U_{2*}$ in a few tens of iterations, which is derived in the following theorem.

\begin{theorem}\label{thm:KKTpoint2}
Under assumptions and conditions of Theorem \ref{thm:equal}, we have
\begin{equation*}
\lim_{t\to\infty} \| \bm U_2^{(t)} - \bm U_{2*} \|_2 = 0.
\end{equation*}
\end{theorem}

Theorem \ref{thm:KKTpoint2} ensures that $\bm\lambda_i^{(t)}$, $\coti$, calculated by Algorithm \ref{alg:MAOM} converge to $\widehat{\bm\lambda}^*$ uniformly as $t \to \infty$.
Therefore, Theorem \ref{thm:convergence} and Theorem \ref{thm:KKTpoint2} together guarantee the following corollary.

\begin{corollary}\label{coro:MAOM}
Under assumptions and conditions of Theorem \ref{thm:equal}, we have
\begin{equation*}
\lim_{t\to\infty} -\sumi \ell_i(\bm\lambda_i^{(t)};\bm\theta_0) \overset{d}{\to} \chi_{(r)}^2,~\text{as}~n\to\infty,
\end{equation*}
where $\bm\lambda_i^{(t)}$ is calculated according to Algorithm \ref{alg:MAOM}.
\end{corollary}

Similarly, we recommend the choice of $\rho = n$ in Algorithm \ref{alg:MAOM}.

\subsection{Linear Convergence Rate for MAOM}\label{Sec:maom-linear}

Comparing with PCM, MAOM halves the number of introduced parameters, and hence simplifies the calculation process and accelerate the convergence speed.
We can further reduce the number of parameters by artificially cutting down the edges in the network, which reduces the redundant penalization terms.
According to the graph theory, the least number for connected graphs with $K$ nodes is $K-1$.
To accomplish it, we can utilize theories of spanning tree, such as Depth-first search algorithm, Breadth-first search algorithm and Shortest path spanning tree algorithm.
This idea is widely adopted to reduce the computation complexity when considering optimization problems with fused lasso penalty \citep{ke2015homogeneity,tang2016fused,zhang2022learning}.
We denote the created spanning tree based on the graph $\mathcal{G} = \{\mathcal{V}, \mathcal{E}\}$ as $\mathcal{G}_{T} = \{ \mathcal{V}, \mathcal{E}_{T} \}$, which has $K$ nodes and $K-1$ edges.
The corresponding incidence matrix is denoted as $\mathbf{A}_{T} \in \mathbb{R}^{(K-1) \times K}$.
Based on the graph $\mathcal{G}_{T}$ and the incidence matrix $\mathbf{A}_{T}$, the linear convergence rate for MAOM can be proved.

\begin{theorem}\label{thm:linear}
Under assumptions and conditions of Theorem \ref{thm:equal}, we have

(a) The incidence matrix $\mathbf{A}_{T}\in \mathbb{R}^{(K-1) \times K}$ is of full row rank;

(b) Based on $\mathcal{G}_{T}$, the convergence rate of Algorithm \ref{alg:MAOM} is linear, that is
\begin{equation*}
\| \bm\Lambda^{(t)} - \widehat{\bm\Lambda}^* \|_2 \le c_0 v^t
\end{equation*}
with some constants $c_0>0$ and $v\in (0,1)$.
\end{theorem}

\section{Simulations}\label{sec:simulation}

In this section, we presented a number of numerical studies to evaluate the finite sample performances of the decentralized distributed empirical likelihood under PCM and MAOM, respectively.
For comparison, we also conducted simulations for the standard empirical likelihood (EL) method \citep{qin1994empirical} with the whole dataset stored in a single machine as the standard.
In Section \ref{sec: sim0}, we provided some basic setups, including a class of particular type of network structures and a slighted modified logarithmic function used in the objective function.
In Section \ref{sec: sim1}, we compared the algorithms with different combinations of $r$ and $p$ in terms of coverage accuracy.
In Section \ref{sec: sim2}, we provided simulations for computational cost of the two algorithms by changing $d$, $n$ and the network structures.
In Section \ref{sec: sim3}, we studied the relationship between the tuning parameter $\rho$ and convergence performances of the two algorithms.

\subsection{Basic Setups}\label{sec: sim0}

\begin{figure}[H]
\centering
\subfigure[$\mathcal{G}_T(10)$]{\includegraphics[width = 3.5cm]{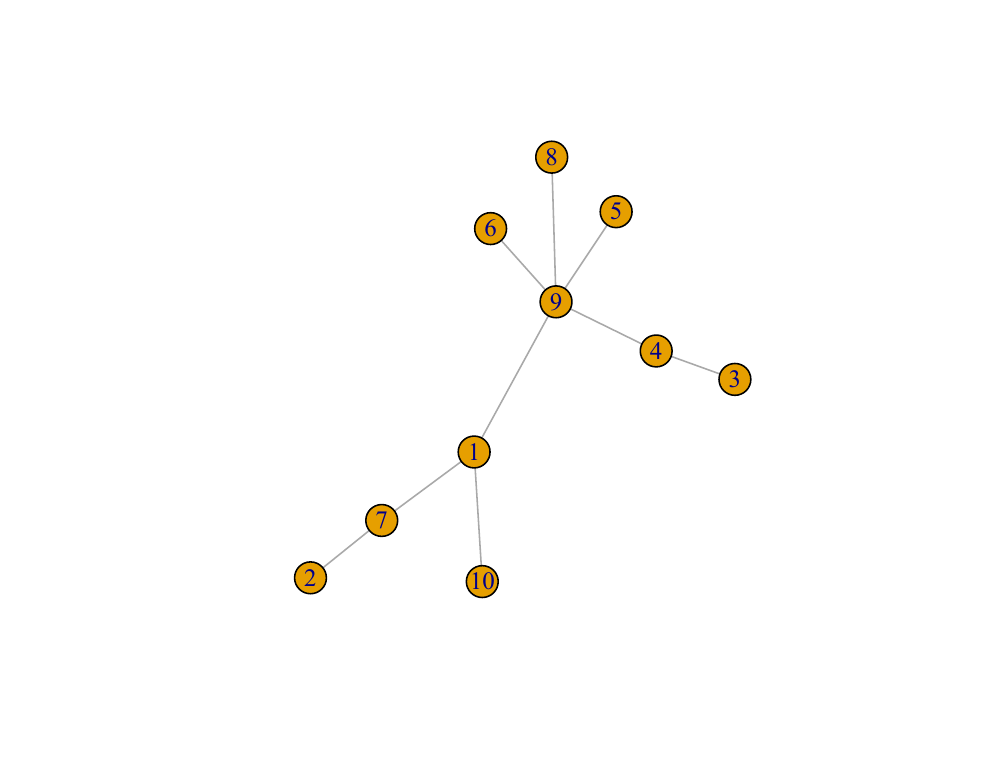}}
\subfigure[$\mathcal{G}(10,0.3)$]{\includegraphics[width = 3.5cm]{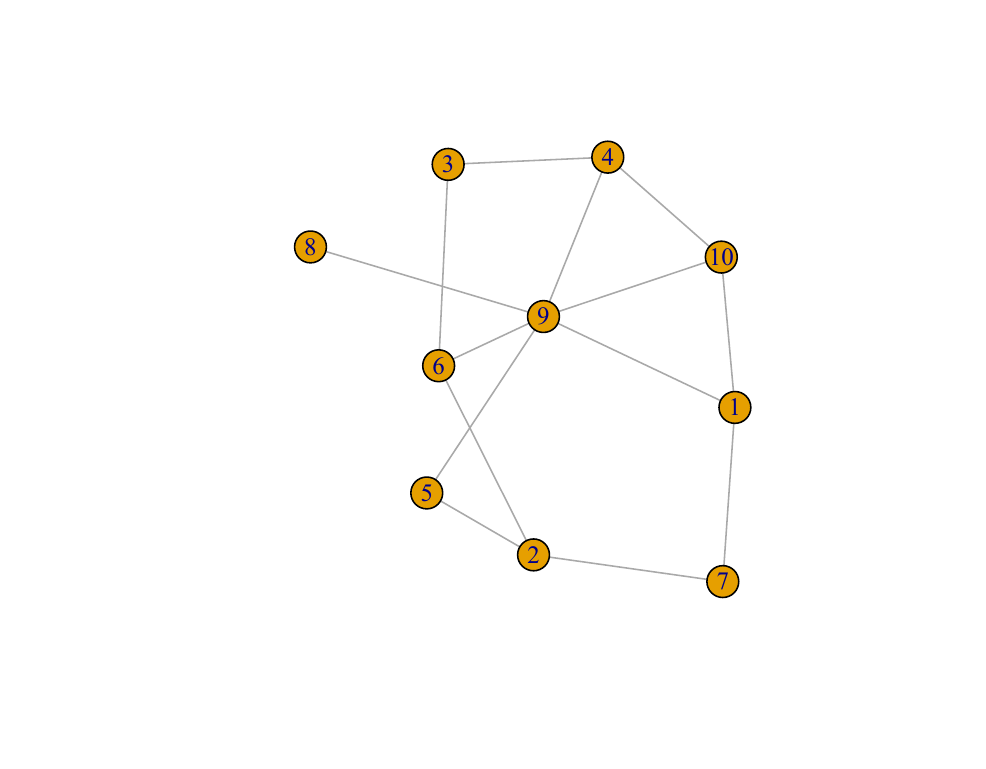}}
\subfigure[$\mathcal{G}(10,0.7)$]{\includegraphics[width = 3.5cm]{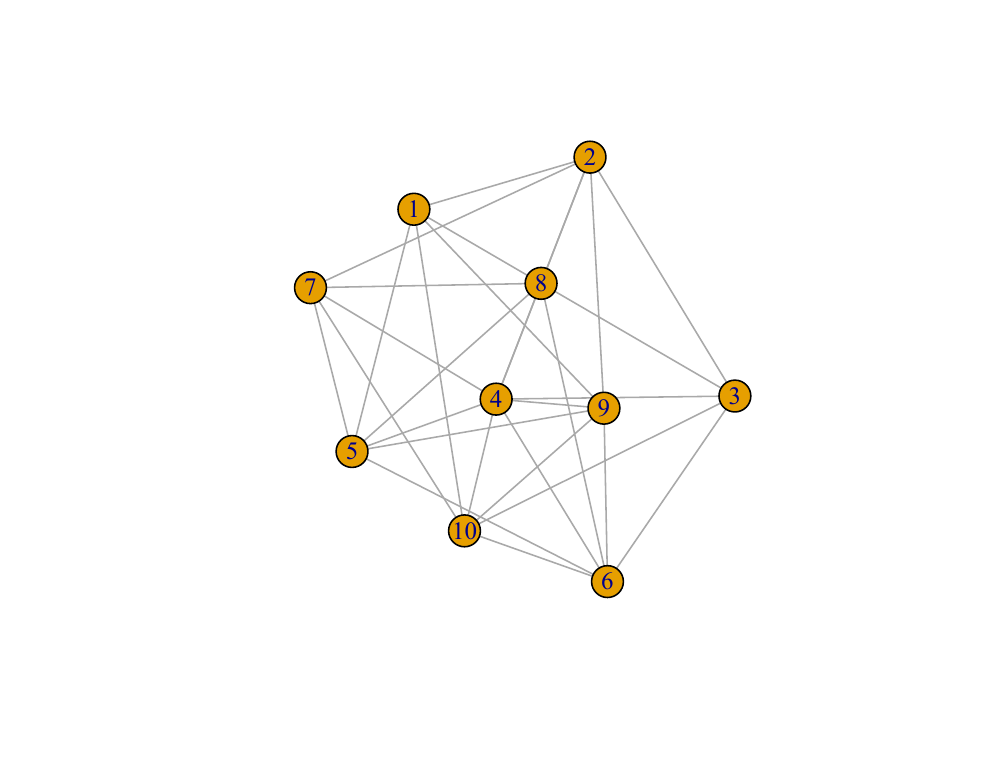}}
\subfigure[$\mathcal{G}(10,1)$]{\includegraphics[width = 3.5cm]{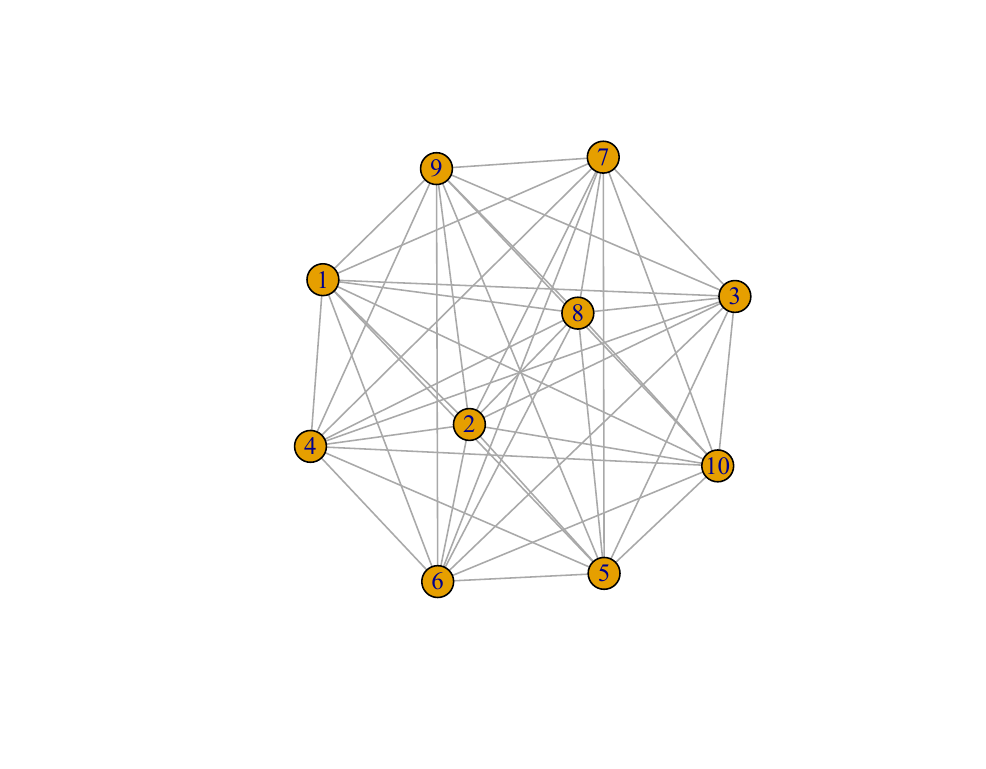}}
\caption{Examples of communication network structures.}\label{sim0_fig}
\end{figure}

In each simulation, once the data are generated, they are then distributed to different nodes randomly.
Meanwhile, the nodes are connected by a specific type of network structure, which is interpreted later.
From the convergence properties of PCM and MAOM, the convergence performances do not rely on the network topology, which is greatly superior to most of existing network algorithms \citep{duchi2012dual,wu2022network}.
Therefore, in our simulations, we recommend Erd\H{o}s–R\'{e}nyi model $\mathcal{G}(K,p_g)$ \citep{erdos1959random}, instead of the elaborate and fragile network topology.
In Erd\H{o}s–R\'{e}nyi model $\mathcal{G}(K,p_g)$, a graph is constructed by connecting the $K$ nodes randomly.
Each potential edge is included in the graph with probability $p_g\in (0,1]$, independently of every other edge.
Hence, the expectation for the number of the edges of graphs generated by $\mathcal{G}(K,p_g)$ is $\oE|\mathcal{E}_{p_g}| = p_g K(K-1)/2$, that is, as $p_g$ increases from $0$ to $1$, the model becomes more and more likely to include graphs with more edges.
According to the conclusion in \cite{erdos1960evolution}, the graph generated by $\mathcal{G}(K,p_g)$ is almost surely connected if $p_g>\log (K)/K$, which guarantees the graphs in the simulations are connected.
To further demonstrate the excellent convergence property of the proposed algorithms, we also generate graphs with spanning tree, which is next denoted as $\mathcal{G}_{T}(K)$.
In Figure \ref{sim0_fig}, we illustrate the graph model by some examples generated by $\mathcal{G}_{T}(K)$ and $\mathcal{G}(K,p_b)$ with $K=10$ and $p_b \in \{0.3, 0.7, 1\}$, respectively.

For stability in implementations, we replace $\ell(\bm\lambda)$ with the following slightly modified objective function:
\begin{equation*}
\ell_*(\bm\lambda) = - 2 \sumi\sumj \log_* \leftsecond 1 + \bm\lambda^\top \bm g(\bm X_{i,j},\bm\theta) \rightsecond,
\end{equation*}
where $\log_*(z)$ is a twice differentiable pseudo-logarithm function with bounded support adopted from \cite{owen2001empirical}:
\begin{equation*}
\log_*(z) = \left\{
\begin{aligned}
&\log(z) & \text{if}~ z\ge \epsilon;\\
&\log(\epsilon) - 1.5 + 2z/\epsilon - z^2/(2\epsilon^2) & \text{if}~ z<\epsilon,
\end{aligned}
\right.
\end{equation*}
where $\epsilon$ is chosen as $N^{-1}$ in our simulations.
Such a modification does not change the empirical log-likelihood ratio statistic theoretically.

\subsection{Empirical Evaluation}\label{sec: sim1}
In this section, we evaluate the decentralized distributed empirical likelihood by the two algorithms with different combinations of $r$ and $p$ in terms of coverage accuracy and interval length.

\textbf{Case 1}: $r=p=1$.
We generated samples from the weibull distribution and calculated the empirical coverages and interval lengths of the $0.05$ quantile of the distribution with confidence level $0.90$ and $0.95$, respectively.
Let $X$ follow weibull distribution with shape parameter being $1.5$ and scale parameter being $200$, and $\beta$ the theoretical $0.05$ quantile.
We conducted simulation with the estimating function $g(X;\beta) = \psi(X-\beta)$, where $\psi(x) = -1$ if $x \le 0$ and $\psi(x) = 0.05/0.95$, otherwise.
$N$ is taken to be $2\times 10^4$, $4\times 10^4$ and $6\times 10^4$ with $K=20,40,60$, respectively. 
In each case, $n=1000$ and the graph was generated by $\mathcal{G}(K,p_g)$ with $p_g = 0.3$.
The simulation experiments were repeated $1000$ times and the results are reported in Table \ref{quantile tab1}.
At each confidence level, the empirical coverage probabilities and the lengths of confidence intervals by the two algorithms PCM and MAOM are close to the ones of the standard EL with the whole dataset.

\begin{table}[hpbt]
\centering
\caption{Empirical coverage probabilities and average length of the confidence intervals constructed by PCM, MAOM and EL with the whole dataset stored in a single machine in the quantile example}\label{quantile tab1}
\begin{tabular}{ccccccc}
\hline
                      &                & \multicolumn{2}{c}{Coverage probabilities} &  & \multicolumn{2}{c}{Interval lengths} \\ \cline{3-4} \cline{6-7} 
                      &                & 90\%                 & 95\%                &  & 90\%              & 95\%             \\ \hline
\multirow{3}{*}{K=20} & PCM            & 0.913                & 0.956               &  & 1.885             & 2.341            \\
                      & MAOM           & 0.913                & 0.956               &  & 1.885             & 2.351            \\
                      & EL             & 0.913                & 0.954               &  & 1.885             & 2.332            \\
                      &                &                      &                     &  &                   &                  \\
\multirow{3}{*}{K=40} & PCM            & 0.907                & 0.947               &  & 1.306             & 1.671            \\
                      & MAOM           & 0.907                & 0.947               &  & 1.306             & 1.682            \\
                      & EL             & 0.907                & 0.947               &  & 1.306             & 1.671            \\
                      &                &                      &                     &  &                   &                  \\
\multirow{3}{*}{K=60} & PCM            & 0.888                & 0.947               &  & 1.056             & 1.380            \\
                      & MAOM           & 0.892                & 0.948               &  & 1.069             & 1.386            \\
                      & EL             & 0.888                & 0.947               &  & 1.056             & 1.380            \\ \hline
\end{tabular}
\end{table}

\textbf{Case 2}: $r=p>1$.
We considered the linear regression model and logistic regression model, respectively.
The linear regression model is given by
\begin{equation*}
Y = \bm X^\top \bm\beta + \varepsilon,
\end{equation*}
where $\bm X \in \mathbb{R}^d$ follows $\mathcal{N}(\bm 0, \mathbf{\Sigma}_{1,d})$ with $\sigma_{ab} = 0.5^{|a-b|}$ for $a,b=1,\ldots,d$ in $\mathbf{\Sigma}_{1,d} = (\sigma_{ab})_{d\times d}$, and $\varepsilon$ follows standard normal distribution.
Let $d=5$ and $\bm\beta = (2,0.5,4,\sqrt{6},-3)^\top$.
Writing $\bm Z=(Y, \bm X^\top)^\top$, we considered the estimating function $\bm g(\bm Z; \bm\beta) = \bm X(Y - \bm X^\top \bm\beta)$.

The logistic regression model is given by 
\begin{equation*}
\Pr(Y=1|\bm X) = \leftsecond 1+ \exp{(\bm X^\top \bm\beta)} \rightsecond^{-1} \exp{(\bm X^\top \bm\beta)},
\end{equation*}
where $\bm X \in \mathbb{R}^d$ follows $\mathcal{N}(\bm 0, \mathbf{\Sigma}_{1,d})$.
Let $d=5$ and $\bm\beta = (1,-2,4,2,-0.5)^\top$.
Writing $\bm Z=(Y, \bm X^\top)^\top$, we considered the estimating function $\bm g(\bm Z; \bm\beta) = \bm X [ Y - \{ 1+ \exp{(-\bm X^\top \bm\beta)} \}^{-1} ]$.
For the two models, $N$ is taken to be $2\times 10^4$, $4\times 10^4$ and $6\times 10^4$ with $K=20,40,60$, respectively. 
In each case, $n=1000$ and the graph was generated by $\mathcal{G}(K,p_g)$ with $p_g = 0.3$.
The simulation experiments were repeated $1000$ times and the results are reported in Table \ref{GLM tab1}.
At each confidence level, the empirical coverage probabilities of the confidence regions of the decentralized distributed EL by the two algorithms and the standard EL with the whole dataset are close to each other.

\begin{table}[hpbt]
\centering
\caption{Empirical coverage probabilities of the confidence regions constructed by PCM, MAOM and EL with the whole dataset stored in a single machine in the generalized linear regression example}\label{GLM tab1}
\begin{tabular}{cccccccccc}
\hline
                          &                & 90\%        & 95\%       &  & 90\%        & 95\%       &  & 90\%        & 95\%       \\ \cline{3-4} \cline{6-7} \cline{9-10} 
Model                     & Method         & \multicolumn{2}{c}{K=20} &  & \multicolumn{2}{c}{K=40} &  & \multicolumn{2}{c}{K=60} \\ \hline
\multirow{3}{*}{Linear}   & PCM            & 0.901       & 0.953      &  & 0.914       & 0.960      &  & 0.897       & 0.950      \\
                          & MAOM           & 0.903       & 0.952      &  & 0.916       & 0.960      &  & 0.909       & 0.952      \\
                          & EL             & 0.902       & 0.952      &  & 0.914       & 0.958      &  & 0.896       & 0.950      \\
                          &                &             &            &  &             &            &  &             &            \\
\multirow{3}{*}{Logistic} & PCM            & 0.906       & 0.958      &  & 0.892       & 0.944      &  & 0.907       & 0.951      \\
                          & MAOM           & 0.905       & 0.958      &  & 0.891       & 0.943      &  & 0.907       & 0.954      \\
                          & EL             & 0.905       & 0.958      &  & 0.890       & 0.943      &  & 0.907       & 0.952      \\ \hline
\end{tabular}
\end{table}

\textbf{Case 3}: $r>p$.
Finally, in order to evaluate the performances of decentralized distributed EL in the case of $r>p$, 
we considered a repeated measures model such that $y_{i,t} = \bm X^\top_{i,t} \bm\beta + \varepsilon_{i,t}$ $(i=1,\ldots,N;\, t = 1,2,3)$, where $\bm X_{i,t}$ follows $\mathcal{N}(\bm 0,\mathbf{\Sigma}_{1,d})$, and $(\varepsilon_{i,1}, \varepsilon_{i,2}, \varepsilon_{i,3})^\top$ follows $\mathcal{N}(\bm 0, \mathbf{\Sigma}_{2,3})$ with $\mathbf{\Sigma}_{2,s} = (\sigma_{ab})_{s\times s}$, $\sigma_{aa} = 1$ for any $a = 1,\ldots, s$ and $\sigma_{ab} = 0.5$ for any $a\neq b$ and $a,b=1,\ldots,s$.
In this simulation, we took $d=2$, $\bm\beta = (1,5)^\top$, and denoted $\bm Y_i = (Y_{i,1},Y_{i,2},Y_{i,3})^\top$ and $\bm X_i = (\bm X_{i,1}^\top, \bm X_{i,2}^\top, \bm X_{i,3}^\top)^\top$ as the response variables and the corresponding predictor variables, respectively.
Writing $\bm Z_i = (\bm Y_i^\top, \bm X_i^\top)^\top$, we conducted simulation with the estimating function $\bm g(\bm Z_i;\bm\beta) = (\bm g_{1}(\bm Z_i;\bm\beta)^\top, \bm g_{2}(\bm Z_i;\bm\beta)^\top)^\top$, where
\begin{equation*}
\bm g_{l}(\bm Z_i;\bm\beta) = \left( \bm X_{i,1}, \bm X_{i,2}, \bm X_{i,3} \right) \mathbf{A}_i^{-1/2} \mathbf{M}_l \mathbf{A}_i^{-1/2} \left(
\begin{aligned}
Y_{i,1} - \bm X^\top_{i,1}\bm\beta \\
Y_{i,2} - \bm X^\top_{i,2}\bm\beta \\
Y_{i,3} - \bm X^\top_{i,3}\bm\beta
\end{aligned}
\right),
\end{equation*}
for $l=1,2$, with $\mathbf{A}_i^{-1/2}$ being the diagonal matrix of the conditional variance, $\mathbf{M}_1 = \mathbf{I}_3$ and $\mathbf{M}_2 = \mathbf{\Sigma}_{2,3}$.
In this setting, the number of estimating equations are larger than the dimension of $\beta$ with $r = 2p$.
We calculated the empirical coverage probabilities of the decentralized distributed EL by PCM and MAOM and the standard EL with the whole dataset with confidence level $0.90$ and $0.95$, respectively.
$N$ is taken to be $2\times 10^4$, $4\times 10^4$ and $6\times 10^4$ with $K=20,40,60$, respectively.
In each case, $n=1000$ and the graph was generated by $\mathcal{G}(K,p_g)$ with $p_g = 0.3$.
The simulation experiments was repeated $1000$ times and the results are reported in Table \ref{repeated tab1}.
At each confidence level, the empirical coverage probabilities of the confidence regions of the decentralized distributed EL by the two algorithms and the standard EL with the whole dataset are close to each other.
\begin{table}[hpbt]
\centering
\caption{Empirical coverage probabilities of the confidence regions constructed by PCM, MAOM and EL with the whole dataset stored in a single machine in the repeated measurements example}\label{repeated tab1}
\begin{tabular}{ccclcclcc}
\hline
               & 90\%        & 95\%       &  & 90\%        & 95\%       &  & 90\%        & 95\%       \\ \cline{2-3} \cline{5-6} \cline{8-9} 
Method         & \multicolumn{2}{c}{K=20} &  & \multicolumn{2}{c}{K=40} &  & \multicolumn{2}{c}{K=60} \\ \hline
PCM            & 0.893       & 0.954      &  & 0.895       & 0.941      &  & 0.900       & 0.940      \\
MAOM           & 0.894       & 0.956      &  & 0.896       & 0.941      &  & 0.901       & 0.939      \\
EL             & 0.901       & 0.957      &  & 0.897       & 0.941      &  & 0.904       & 0.943      \\ \hline
\end{tabular}
\end{table}

\subsection{Computational Cost}\label{sec: sim2}
In the last subsection, the decentralized distributed EL by PCM and MAOM has almost same performances as the standard EL with the whole dataset in terms of coverage accuracy and interval length. 
In this section, we evaluate the performances of the two algorithms for different combinations $d$, $N$ and $K$ in terms of computational cost, respectively.
We considered the inference problem for mean, and generate samples as independent copies of $\bm X = (X_1,\ldots, X_d)^\top \sim \mathcal{N}(\bm\mu_{1,d}, \mathbf{\Sigma}_{2,d})$, where $\bm\mu_{1,d} = (1,\ldots,1)^\top$.
We considered the estimating function $\bm g(\bm X;\bm\mu) = \bm X - \bm\mu$.
We took different combinations of $d\in \{3,5\}$, $N \in\{2.5\times 10^5,5\times 10^5\}$ and $K=50$, and generated the network structure by $\mathcal{G}_{T}(K)$ and $\mathcal{G}(K,p_g)$ with $p_g \in\{0.1,0.3, 0.5,0.7, 1\}$.The simulation experiments were repeated 50 times.
The results are listed in Table \ref{sim2 tab1}, including the average number of iterations and the average CPU time cost (in seconds) for the two algorithms when computing the corresponding Lagrange multipliers.
The R programming language \citep{r2023r} is used to implement each method in Windows 10 (2x Intel Xeon Gold 5220R CPU 2.20GHz) with 48 cores and 128 GB memory.

\begin{table}[hpbt]
\centering
\caption{The average number of iterations (denoted as Iter-$\cdot$) and the average CPU time cost (in seconds, denoted as Time-$\cdot$) needed for convergence of PCM and MAOM.}\label{sim2 tab1}
\begin{tabular}{clrrrrrr}
\hline
$(d,n)$                      &           & $\mathcal{G}_{T}(50)$ & $\mathcal{G}(50,0.1)$ & $\mathcal{G}(50,0.3)$ & $\mathcal{G}(50,0.5)$ & $\mathcal{G}(50,0.7)$ & $\mathcal{G}(50,1)$ \\ \hline
\multirow{4}{*}{$(3,5000)$}  & Iter-PCM  & 44.6                & 38.4                  & 39.9                  & 51.0                  & 59.1                  & 66.2                \\
                             & Time-PCM  & 4.6                 & 6.9                   & 7.9                   & 11.6                  & 15.9                  & 21.0                \\
                             & Iter-MAOM & 26.3                & 36.6                  & 55.0                  & 64.8                  & 68.7                  & 68.8                \\
                             & Time-MAOM & 1.7                 & 2.6                   & 3.9                   & 4.7                   & 4.8                   & 4.4                 \\
                             &           &                     &                       &                       &                       &                       &                     \\
\multirow{4}{*}{$(3,10000)$} & Iter-PCM  & 47.8                & 40.8                  & 44.5                  & 58.0                  & 69.0                  & 80.1                \\
                             & Time-PCM  & 7.6                 & 10.6                  & 11.6                  & 17.2                  & 23.6                  & 30.2                \\
                             & Iter-MAOM & 27.9                & 39.3                  & 63.2                  & 78.0                  & 86.7                  & 91.3                \\
                             & Time-MAOM & 3.4                 & 5.5                   & 8.9                   & 11.1                  & 11.6                  & 11.5                \\
                             &           &                     &                       &                       &                       &                       &                     \\
\multirow{4}{*}{$(5,5000)$}  & Iter-PCM  & 98.7                & 64.0                  & 83.6                  & 99.5                  & 109.8                 & 127.1               \\
                             & Time-PCM  & 26.4                & 28.6                  & 41.5                  & 60.8                  & 77.1                  & 107.8               \\
                             & Iter-MAOM & 36.2                & 41.0                  & 72.5                  & 92.4                  & 105.8                 & 116.4               \\
                             & Time-MAOM & 2.3                 & 2.9                   & 5.4                   & 6.8                   & 7.6                   & 7.7                 \\
                             &           &                     &                       &                       &                       &                       &                     \\
\multirow{4}{*}{$(5,10000)$} & Iter-PCM  & 131.9               & 135.6                 & 129.1                 & 139.3                 & 139.2                 & 165.3               \\
                             & Time-PCM  & 55.7                & 74.2                  & 87.9                  & 112.3                 & 126.3                 & 174.0               \\
                             & Iter-MAOM & 38.5                & 44.6                  & 84.9                  & 112.4                 & 133.3                 & 153.8               \\
                             & Time-MAOM & 4.8                 & 6.4                   & 12.2                  & 16.2                  & 18.6                  & 20.0                \\ \hline
\end{tabular}
\end{table}

Table \ref{sim2 tab1} demonstrates the superiority of MAOM over PCM.
Generally, the average number of iterations and CPU time cost needed for convergence of MAOM is much smaller than that of PCM.
Furthermore, MAOM requires significantly less time than PCM during each iteration.
It is worth noting that the average number of iterations needed for MAOM increases as the number of edges increases.
The fact coincides with Theorem \ref{thm:linear} that MAOM with spanning tree structure enjoys the fastest convergence speed.
However, the average number of iterations needed for PCM does not exhibit a clear trend as the number of edge increases.
This behavior may be attributed to the introduction of superabundant parameters.
From the results obtained with different $d$, we observe that the average number of iterations needed for PCM increases rapidly as $d$ increases, while that for MAOM remains relatively stable.
The simulation result meets our analysis.
In PCM, we solve the optimization problem related to $\sumi \ell_i(\bm\lambda_i)$ by numerical calculation, which costs lot of time.
In MAOM, we solve the modified approximation problem by computing the closed form vector, which costs much less time.

\subsection{Selection of the Tuning Parameter}\label{sec: sim3}

\begin{figure}[H]
\centering
\subfigure[$n = 1000$]{\includegraphics[width = 7cm]{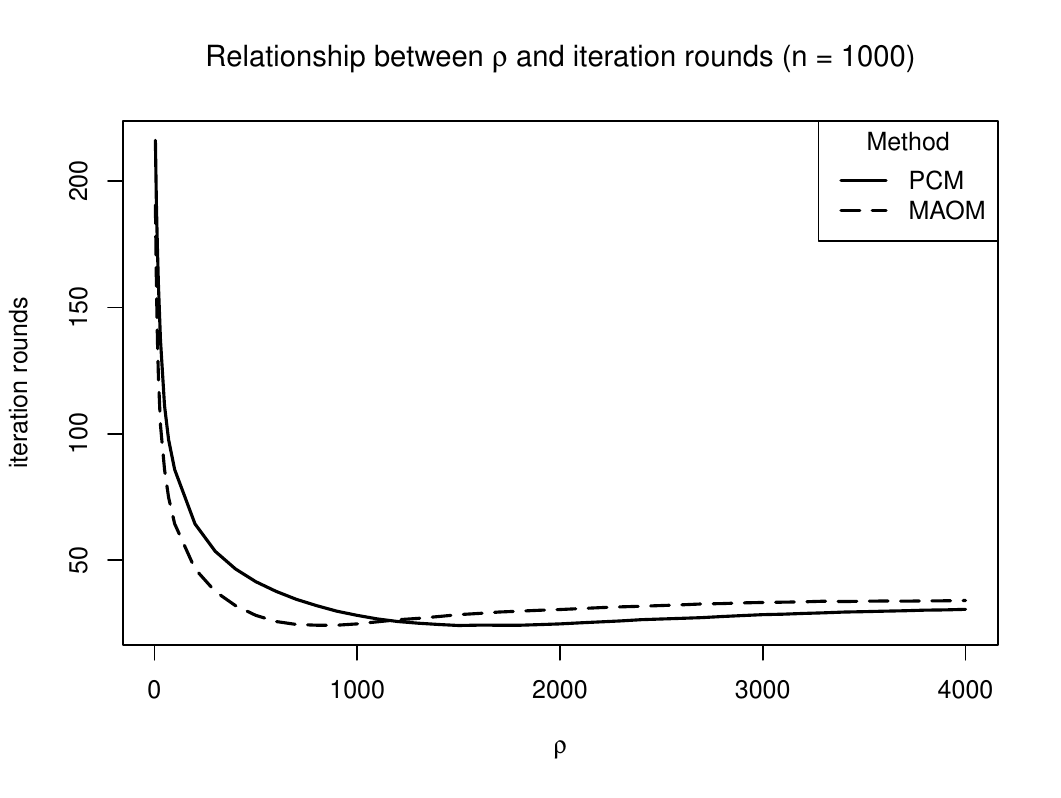}}
\subfigure[$n = 5000$]{\includegraphics[width = 7cm]{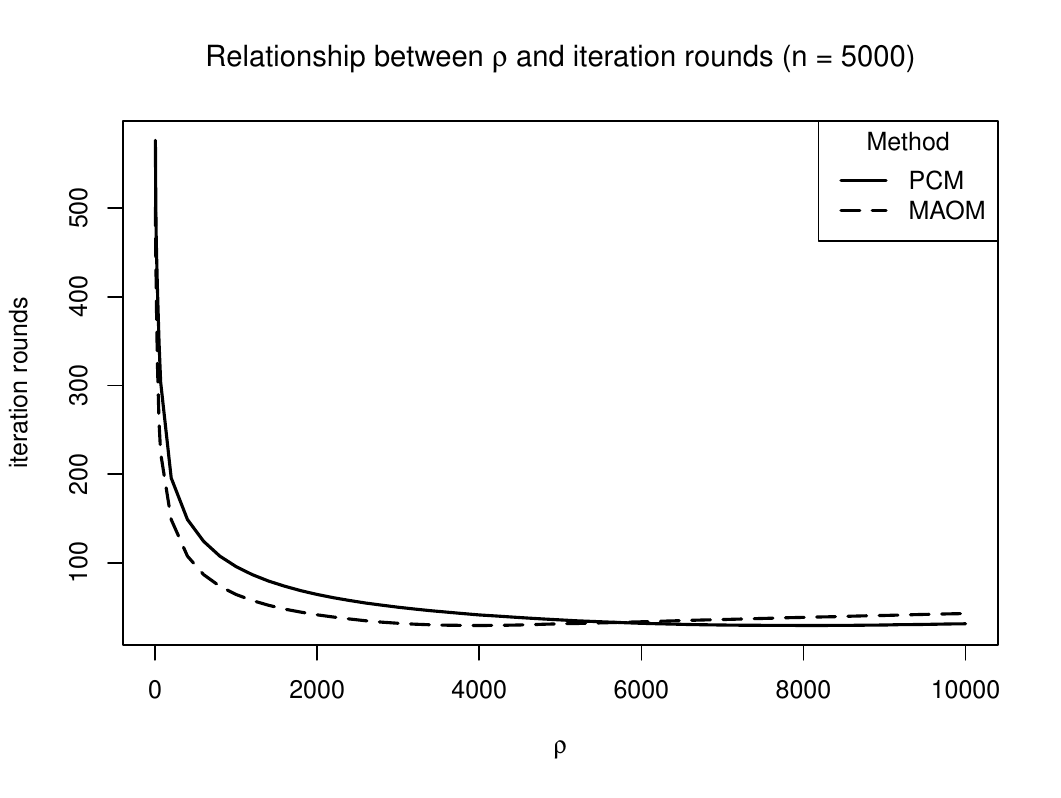}}
\caption{The relationship between the tuning parameter $\rho$ and the average number of iteration rounds needed for convergence of the two algorithms. We take $n = 1,000$ and $n = 5,000$ for (a) and (b), respectively.}\label{rho fig}
\end{figure}

There is only a single parameter $\rho$ in both PCM and MAOM, and it is demonstrated that both of the two algorithms converge for any $\rho>0$.
Nevertheless, $\rho$ has a direct impact on the convergence rate of the algorithms.
An arbitrarily selected $\rho$ can result in an awfully slow convergence speed.
There existing several literature concerning the optimal parameter selection for ADMM.
However, the existing literature mostly focus on quadratic problems \citep{raghunathan2014alternating,ghadimi2015optimal}.
For the problems with complex construction, the direct analysis for the convergence rate is no longer available.
For the proposed two methods, we recommend the choice of $\rho = O(n)$, and we take $\rho = n$ in our simulations above.
Intuitively, the optimal tuning parameter can accelerate the convergence speed by facilitating that the elements of $\bm U_{s}^{(t)}$ approach the ones of $\bm U_{s*}$ for $s=1,2$, respectively.
Based on the fact that $\widehat{\bm\lambda}^* = O_p(n^{-1/2})$, $\bm c_{ii'}^\iterO$ in PCM and $\bm z_{ii'}^\iterO$ in MAOM are also $O_p(n^{-1/2})$ for $(i,i')$ with $e_{ii'} \in \mathcal{E}$ and $t\ge 1$.
Meanwhile, from the update strategies for $\bm\Lambda^\iterO$, we know that $\bm v_{ii'}^\iterO$ in PCM and $\bm t_{ii'}^\iterO$ in MAOM are $O_p(n^{1/2})$.
Therefore, the choice of $\rho=O(n)$ is suitable.
To illustrate it, we provided an example to show the relationship between the tuning parameter $\rho$ and the number of iteration rounds needed for convergence.
Considering the inference problem for mean, we generated samples as independent copies of $\bm X = (X_1,\ldots, X_d)^\top \sim \mathcal{N}(\bm\mu_{1,d}, \mathbf{\Sigma}_{2,d})$, where $\bm\mu_{1,d} = (1,\ldots,1)^\top$.
In this section, let $d=3$, $K = 20$, and the graph be generated by $\mathcal{G}(K,p_g)$ with $p_g = 0.2$.
We considered the estimating function $\bm g(\bm X;\bm\mu) = \bm X - \bm\mu$.
Let $n \in \{1000,5000\}$.
We recorded the average number of iteration rounds needed for convergence by the two algorithms with different tuning parameter $\rho$.
The simulation experiments were repeated 50 times.
The results for PCM and MAOM are plotted in Figure \ref{rho fig}.
The results and our analysis coincide that the choice of $\rho = n$ accelerates the convergence speed of the two algorithms.

\section{Real Data Analysis}\label{sec:real data}

\subsection{Analysis of Census Income Dataset}
In this section, we apply the proposed methods to a census income dataset \citep{kohavi1996census}, which is extracted from the 1994 Census database.
The dataset contains $48,842$ observations, which is available on \url{https://archive.ics.uci.edu/dataset/20/census+income}.
The response variable is whether a person’s income exceeds \$$50$K a year, and we consider a logistic regression model based on the covariates: $x_1$, age; $x_2$, final weight (Fnlwgt); $x_3$, highest level of education in numerical form; $x_4$, capital change (capital gain - capital loss); $x_5$, hours worked per week.
The values $x_2$ are assigned by Population Division at the Census Bureau, and they are related to the socio-economic characteristic, that is, people with similar socio-economic characteristics have similar weights.
The values $x_5$ are the gain or loss in income due to investments.

To eliminate the effect of scale, we standardized all the variables.
We denote the parameter corresponding to $x_i$ as $\beta_i$ for $i = 1,\ldots, 5$, and denote the intercept parameter by $\beta_0$.
We equally divide the observations to $K=20$ nodes randomly, and the graph is generated by $\mathcal{G}(20, 0.3)$.
We calculated $95\%$ confidence interval of every component $\beta_i$ for $i=1,\ldots,5$, with Lagrange multipliers calculated by PCM and MAOM, respectively.
The results are displayed in Table \ref{real1 tab1}.
For comparison, we also calculated their $95\%$ confidence intervals with Lagrange multipliers calculated by EL with the whole dataset stored in a single machine.
Table \ref{real1 tab1} shows the confidence intervals calculated by the proposed decentralized distributed algorithms are similar to the ones calculated by the standard EL with the whole dataset stored in a single machine.
Furthermore, the result and our perceptual intuition coincide.
That is, all the covariates have positive effect on the income.
Higher work time and higher education level facilitate higher income.

\begin{table}[hpbt]
\centering
\caption{95\% confidence intervals for parameters in census income model estimated by PCM, MAOM and EL with the whole dataset stored in a single machine}\label{real1 tab1}
\begin{tabular}{ccccccc}
\hline
                    &  & PCM              &  & MAOM             &  & EL    \\ \cline{3-3} \cline{5-5} \cline{7-7} 
$\widehat{\beta}_0$ &  & (-2.017, -1.957) &  & (-2.016, -1.958) &  & (-2.016, -1.958) \\
$\widehat{\beta}_1$ &  & (0.485, 0.534)   &  & (0.485, 0.533)   &  & (0.485, 0.533)   \\
$\widehat{\beta}_2$ &  & (0.018, 0.069)   &  & (0.018, 0.069)   &  & (0.018, 0.069)   \\
$\widehat{\beta}_3$ &  & (0.683, 0.742)   &  & (0.684, 0.741)   &  & (0.684, 0.740)   \\
$\widehat{\beta}_4$ &  & (0.386, 0.437)   &  & (0.386, 0.436)   &  & (0.386, 0.436)   \\
$\widehat{\beta}_5$ &  & (0.173, 0.265)   &  & (0.174, 0.264)   &  & (0.174, 0.265)   \\ \hline
\end{tabular}
\end{table}

\subsection{Ford Gobike Data Analysis}

In this section, we apply the proposed methods to the dataset of Ford gobike trip data.
This dataset is taken from \url{https://www.kaggle.com/datasets/chirag02/ford-gobike-2019feb-tripdata}, which represents trips taken by members of the service for the month of February of 2019.
The full dataset contains 183,412 records and consists of information about trips taken by service's members, their types, their age, time of starting and ending trips, duration of trips, and so on.
We are interested in the influence factors for the member's type (subscriber or customer), including duration of trips, time of starting trips, and age.
For analysis, we consider a logistic regression model, where the response variable denotes whether the member is a subscriber (binary variable taken 1 if the member is a subscriber and 0 otherwise).
We take covariates: $x_1$, the duration of trips; $x_2$, starting trip on the weekend/workday (binary variable taken 1 if the trip starts on weekend and 0 otherwise); $x_3$, starting trip for commuting or not (binary variable taken 1 if the trip starts during 7 a.m. to 10 a.m. or 5 p.m. to 8 p.m. and 0 otherwise); $x_4$, age.
After dropping the missing samples and abnormal samples (such as members whose ages are larger than 100), we have 174,440 data points.

To eliminate the effect of scale, we standardized all the variables.
We denote the parameter corresponding to $x_i$ as $\beta_i$ for $i = 1,\ldots, 4$, and denote the intercept parameter by $\beta_0$.
We equally divide the observations to $K=20$ nodes randomly, the graph is generated by $\mathcal{G}(20, 0.3)$.
We estimate $95\%$ confidence interval for each component $\beta_i$ for $i=0,\ldots,4$, with Lagrange multipliers calculated by PCM and MAOM, respectively.
The result is displayed in Table \ref{real2 tab1}.
For comparison, we also estimate their $95\%$ confidence intervals with Lagrange multipliers calculated by EL with the whole dataset stored in a single machine.
Table \ref{real2 tab1} shows the confidence intervals calculated by the proposed decentralized distributed algorithms are similar to the ones calculated by the standard EL with the whole dataset stored in a single machine.
Furthermore, the result and our perceptual intuition coincide.
That is, $\widehat{\beta}_1<0$ implies that customers always ride for a longer time for each trip because they pay in each trip, while subscribers always ride for a shorter time.
$\widehat{\beta}_2<0$ together with $\widehat{\beta}_3>0$ implies that members always subscribe for regular commuting in the workday.
Compared with other influence factors, the age influences less.

\begin{table}[hpbt]
\centering
\caption{95\% confidence intervals for parameters in gobike model estimated by PCM, MAOM and EL with the whole dataset stored in a single machine}\label{real2 tab1}
\begin{tabular}{ccccccc}
\hline
                    &  & PCM              &  & MAOM             &  & EL       \\ \cline{3-3} \cline{5-5} \cline{7-7} 
$\widehat{\beta}_0$ &  & (2.365, 2.410)   &  & (2.360, 2.410)   &  & (2.360, 2.410)   \\
$\widehat{\beta}_1$ &  & (-0.474, -0.446) &  & (-0.474, -0.446) &  & (-0.474, -0.446) \\
$\widehat{\beta}_2$ &  & (-0.537, -0.455) &  & (-0.537, -0.457) &  & (-0.537, -0.457) \\
$\widehat{\beta}_3$ &  & (0.140, 0.208)   &  & (0.140, 0.205)   &  & (0.140, 0.205)   \\
$\widehat{\beta}_4$ &  & (0.052, 0.084)   &  & (0.052, 0.084)   &  & (0.052, 0.084)   \\ \hline
\end{tabular}
\end{table}

\section{Discussion}\label{discussion}

In this article, we propose an optimization problem concerning empirical likelihood inference over decentralized networks and establish the asymptotic properties of the distributed empirical log-likelihood ratio statistic.
Additionally, we propose two algorithms, PCM and MAOM, to solve the optimization problem.
It is worth mentioning that the two algorithms are not constrained by the network structure, indicating their universality across different connected network.
The convergence rate of PCM is hindered by an excessive number of parameters although it has clear intuition and employs a novel decoupling technique.
To address it, we propose MAOM, which significantly 
reduces computation burden and accelerates computation speed.
The specially designed regularization function in MAOM enhances the convexity of the approximation function and decouples the penalty term.
Furthermore, for any connected graph, we can figure out a spanning tree that maintains connectivity with the fewest number of edges.
Based on the spanning tree, MAOM enjoys a linear convergence rate.
Our numerical experiments demonstrate that the algorithms converge in a few tens of iterations, significantly outperforming the majority of existing decentralized algorithms.
Meanwhile, the performances of the decentralized distributed algorithms is comparable to that based on the whole dataset.
The ideas presented in this article can be easily extended to the problems of M-estimation over decentralized networks.

\begin{appendix}
\section*{Proof of the main results}

\subsection{Proof of Theorem 2.1}
Recall the definition of $\ell_{lasso}(\bm\Lambda)$:
\begin{equation*}
\ell_{lasso}(\bm\Lambda) = -2\sumi\sumj \log \leftsecond 1+ \bm\lambda_i^\top \bm g(\bm X_{i,j}, \bm\theta) \rightsecond + \sumi \sum_{\stackrel{i'\in \mathcal{N}_i}{i'>i}} \eta_n \| \bm\lambda_i - \bm\lambda_{i'}\|_2.
\end{equation*}
To prove Theorem 2.1, it is sufficient to show $\widehat{\bm\Lambda}^*$ is the strictly global minimizer of $\ell_{lasso}(\bm\Lambda)$ with probability tending to $1$ under the assumptions.
Given $\varepsilon_n \ge 0$, define
\begin{equation*}
\mathcal{T}_{\varepsilon_n}^i := \left\{ \bm\lambda:\, 1 + \bm\lambda^\top \bm g(\bm X_{i,j},\bm\theta) \ge \varepsilon_n ~\text{for all}~ \cotj \right\},
\end{equation*}
\begin{equation*}
\widebar{\mathcal{T}}_{\varepsilon_n} := \bigcap_{i=1}^K \mathcal{T}_{\varepsilon_n}^i = \left\{ \bm\lambda:\, 1 + \bm\lambda^\top \bm g(\bm X_{i,j},\bm\theta) \ge \varepsilon_n ~\text{for all}~ \coti, ~\text{and}~ \cotj \right\},
\end{equation*}
and
\begin{equation*}
\begin{aligned}
\mathcal{D}_{\varepsilon_n} := {}&{} \left\{ \bm\Lambda = (\bm\lambda_1^\top,\ldots, \bm\lambda_K^\top)^\top:\, \bm\lambda_i \in \mathcal{T}_{\varepsilon_n}^i ~\text{for}~ \coti \right\},\\
\widebar{\mathcal{D}}_{\varepsilon_n} := {}&{} \left\{ \bm\Lambda = (\bm\lambda_1^\top,\ldots, \bm\lambda_K^\top)^\top:\, \bm\lambda_i \in \widebar{\mathcal{T}}_{\varepsilon_n} ~\text{for}~ \coti \right\}.
\end{aligned}
\end{equation*}
From \cite{owen2001empirical}, although the domain of $\log(x)$ is $\{x:\, x>0\}$, the domain for the Lagrange multiplier is slightly tighter because the probability placed on each value is not bigger than $1$.
Hence, the domain of $\bm\Lambda$ is $\mathcal{D}_{n^{-1}}$, and the domain of $\widehat{\bm\Lambda}^*$ is $\widebar{\mathcal{D}}_{N^{-1}}$.
We next prove $\ell_{lasso}(\widehat{\bm\Lambda}^*) < \ell_{lasso}(\bm\Lambda)$ for $\forall \bm\Lambda \in \mathcal{D}_{N^{-1}}$, which is sufficient to prove Theorem 2.1 because $\mathcal{D}_{n^{-1}} \subset \mathcal{D}_{N^{-1}}$.

First, under the condition $\bm\theta \in \mathbb{R}^p$, $\eta_n/N^2 \to \infty$, we show $\Pr (\widehat{\bm\Lambda} = \widehat{\bm\Lambda}^*) \to 1$ as $n\to\infty$.
In this case, we complete the proof by analyzing (i) $\bm\Lambda \in \widebar{\mathcal{D}}_{N^{-1}}$ and (ii) $\bm\Lambda \in \mathcal{D}_{N^{-1}} \setminus \widebar{\mathcal{D}}_{N^{-1}}$.

(i) $\bm\Lambda \in \widebar{\mathcal{D}}_{N^{-1}}$:
For any $\bm\Lambda = (\bm\lambda_1^\top,\ldots,\bm\lambda_K^\top)^\top \in \widebar{\mathcal{D}}_{N^{-1}}$, define $\widebar{\bm\Lambda} = (\widebar{\bm\lambda}^\top, \ldots, \widebar{\bm\lambda}^\top)^\top$, where $\widebar{\bm\lambda} = K^{-1} \sumi \bm\lambda_i$.
It is easy to see that $\widebar{\bm\Lambda}$ is well-defined because $\widebar{\bm\lambda} \in \widebar{\mathcal{T}}_{N^{-1}}$ and $\widebar{\bm\Lambda} \in \widebar{\mathcal{D}}_{N^{-1}}$.
We complete the proof by ordering the value of $\ell_{lasso}(\widehat{\bm\Lambda}^*)$, $\ell_{lasso}(\widebar{\bm\Lambda})$ and $\ell_{lasso}(\bm\Lambda)$.

Recalling the definition of $\widehat{\bm\Lambda}^*$, it follows
\begin{equation*}
\ell_{lasso}(\widehat{\bm\Lambda}^*) = \ell(\widehat{\bm\lambda}^*) \le \ell(\widebar{\bm\lambda}) = \ell_{lasso}(\widebar{\bm\Lambda})
\end{equation*}
because $\ell(\bm\lambda)$ is a strictly convex function.
Hence, it holds
\begin{equation}\label{P.Orcal.i.1}
\ell_{lasso}(\widehat{\bm\Lambda}^*) < \ell_{lasso}(\widebar{\bm\Lambda})
\end{equation}
for $\forall \bm\Lambda \in \widebar{\mathcal{D}}_{N^{-1}}$ with $\widebar{\bm\Lambda} \neq \widehat{\bm\Lambda}^*$.

From the definition of $\ell_{lasso}(\widebar{\bm\Lambda})$ and $\ell_{lasso}(\bm\Lambda)$, it holds
\begin{equation}\label{P.Oracle.i.split}
\ell_{lasso}(\bm\Lambda) - \ell_{lasso}(\widebar{\bm\Lambda}) = A_1 + A_2,
\end{equation}
where
\begin{equation*}
A_1 = - 2\sumi (\bm\lambda_i - \widebar{\bm\lambda} )^\top \sumj \frac{\bm g(\bm X_{i,j},\bm\theta)}{ 1 + \widecheck{\bm\lambda}_i^\top \bm g(\bm X_{i,j},\bm\theta)}
\end{equation*}
with $\widecheck{\bm\lambda}_i = a_i \bm\lambda_i + (1-a_i) \widebar{\bm\lambda}$ for some $a_i \in (0,1)$, $\coti$, according to Taylor's expansion, and
\begin{equation*}
A_2 = \eta_n \sumi \sum_{\stackrel{i'\in \mathcal{N}_i}{i'>i}} \| \bm\lambda_i - \bm\lambda_{i'}\|_2.
\end{equation*}

For the first term, take
\begin{equation*}
\bm G_i = \frac{1}{n} \sumj \frac{\bm g(\bm X_{i,j},\bm\theta)}{ 1 + \widecheck{\bm\lambda}_i^\top \bm g(\bm X_{i,j},\bm\theta)} ~\text{for}~ \coti.
\end{equation*}
We have $A_1 = -2n \sumi \bm G_i^\top (\bm\lambda_i - \widebar{\bm\lambda})$.
It is easy to see $\widecheck{\bm\lambda}_i \in \widebar{\mathcal{T}}_{N^{-1}}$ for all $\coti$.
Hence, it holds
\begin{equation}\label{P.Oracle.i.GQ}
\Pr \left( N^{-1} \| \bm G_i \|_2 < \| \bm Q_i \|_2 \right) \to 1, ~\text{as}~n\to\infty,
\end{equation}
with $\bm Q_i = n^{-1} \sumj \bm g( \bm X_{i,j}, \bm\theta)$ being asymptotically normal distributed.
It can be proved that $\max_{\coti} \|\bm Q_i\|_2 = O_p(1)$, which implies $\max_{\coti} \|\bm G_i\|_2 = O_p(N)$ according to \eqref{P.Oracle.i.GQ}.
Therefore, we have
\begin{equation*}
|A_1| \le nK \max_{\coti} \|\bm G_i\|_2 \max_{\coti} \| \bm\lambda_i - \widebar{\bm\lambda} \|_2 = \max_{\coti} \| \bm\lambda_i - \widebar{\bm\lambda} \|_2 O_p(N^2)
\end{equation*}
for $\bm\Lambda \in \widebar{\mathcal{D}}_{N^{-1}}$.
Under the condition $\eta_n/N^2 \to \infty$, we have
\begin{equation}\label{P.Oracle.i.A1}
|A_1| = \max_{\coti} \| \bm\lambda_i - \widebar{\bm\lambda} \|_2 o_p(\eta_n)
\end{equation}
for $\bm\Lambda \in \widebar{\mathcal{D}}_{N^{-1}}$.

For the second term, it holds
\begin{equation}\label{P.Oracle.i.Path}
\sumi \sum_{\stackrel{i'\in \mathcal{N}_i}{i'>i}} \| \bm\lambda_i - \bm\lambda_{i'}\|_2 \ge \max_{i,i'=1,\ldots,K} \| \bm\lambda_i - \bm\lambda_{i'} \|_2
\end{equation}
because there exists a path which connects any $i\neq i'$ for $i,i' =1,\ldots,K$, according to the connectedness of the graph $\mathcal{G}$.
From the definition of $\widebar{\bm\lambda}$, we have
\begin{equation}\label{P.Oracle.i.inequation}
\begin{aligned}
\max_{\coti} \|\bm\lambda_i - \widebar{\bm\lambda} \|_2 = {}&{} \max_{\coti} \|K^{-1} \sum_{i'=1}^K (\bm\lambda_i - \bm\lambda_{i'}) \|_2\\
\le {}&{} \max_{\coti} K^{-1}\sum_{i'=1}^K \|\bm\lambda_i - \bm\lambda_{i'} \|_2\\
\le {}&{} \max_{i,i'=1,\ldots,K} \| \bm\lambda_i - \bm\lambda_{i'} \|_2.
\end{aligned}
\end{equation}
From the definition of $A_2$, combining \eqref{P.Oracle.i.Path} and \eqref{P.Oracle.i.inequation}, we have
\begin{equation}\label{P.Oracle.i.A2}
A_2 \ge \eta_n \max_{\coti} \|\bm\lambda_i - \widebar{\bm\lambda} \|_2.
\end{equation}
Combining \eqref{P.Oracle.i.split}, \eqref{P.Oracle.i.A1} and \eqref{P.Oracle.i.A2}, it holds
\begin{equation*}
\ell_{lasso}(\bm\Lambda) - \ell_{lasso}(\widebar{\bm\Lambda}) = A_1 + A_2 \ge \leftsecond \eta_n + o_p(\eta_n) \rightsecond \max_{i,i'=1,\ldots,K} \| \bm\lambda_i - \bm\lambda_{i'} \|_2 > 0,
\end{equation*}
which implies
\begin{equation}\label{P.Orcal.i.2}
\ell_{lasso}(\widebar{\bm\Lambda}) < \ell_{lasso}(\bm\Lambda)
\end{equation}
with probability tending to $1$, for $\forall \bm\Lambda \in \widebar{\mathcal{D}}_{N^{-1}}$ with $\bm\Lambda \neq \widebar{\bm\Lambda}$.

Combining \eqref{P.Orcal.i.1} and \eqref{P.Orcal.i.2}, with probability tending to $1$, we have $\ell_{lasso}(\widehat{\bm\Lambda}^*) < \ell_{lasso}(\bm\Lambda)$ for $\forall \bm\Lambda \in \widebar{\mathcal{D}}_{N^{-1}}$ and $\bm\Lambda \neq \widehat{\bm\Lambda}^*$ since either the event $\mathcal{A} := \{ \bm\Lambda \neq \widebar{\bm\Lambda} \}$ or the event $\mathcal{B} := \{ \widebar{\bm\Lambda} \neq \widehat{\bm\Lambda}^* \}$ holds.

(ii) $\bm\Lambda \in \mathcal{D}_{N^{-1}} \setminus \widebar{\mathcal{D}}_{N^{-1}}$:
For each given $\bm\Lambda \in \mathcal{D}_{N^{-1}} \setminus \widebar{\mathcal{D}}_{N^{-1}}$, we complete the proof by constructing a corresponding intermediate value $\widetilde{\bm\Lambda} \in \widebar{\mathcal{D}}_{N^{-1}}$, which satisfies $\ell_{lasso}(\widetilde{\bm\Lambda}) < \ell_{lasso}(\bm\Lambda)$.

Without loss of generality, for $\bm\Lambda = (\bm\lambda_1^\top,\ldots, \bm\lambda_K^\top)^\top \in \mathcal{D}_{N^{-1}} \setminus \widebar{\mathcal{D}}_{N^{-1}}$, we assume $\bm\lambda_1 \in \mathcal{T}_{N^{-1}}^1 \setminus \widebar{\mathcal{T}}_{N^{-1}}$ and $\bm\lambda_i \in \widebar{\mathcal{T}}_{N^{-1}}$ for $i=2,\ldots,K$.
Let $\widetilde{\bm\lambda}_1 := \arg\min_{\bm\lambda \in \widebar{\mathcal{T}}_{N^{-1}}} \sum_{i' \in \mathcal{N}_1} \| \bm\lambda - \bm\lambda_{i'} \|_2$ and $\widetilde{\bm\Lambda} = (\widetilde{\bm\lambda}_1^\top, \bm\lambda_2^\top, \ldots, \bm\lambda_K^\top) \in \widebar{\mathcal{D}}_{N^{-1}}$.
From the definition of $\ell_{lasso}(\bm\Lambda)$ and $\ell_{lasso}(\widetilde{\bm\Lambda})$, similar to \eqref{P.Oracle.i.split}, we have
\begin{equation}\label{P.Oracle.ii.split}
\ell_{lasso}(\bm\Lambda) - \ell_{lasso}(\widetilde{\bm\Lambda}) = \widetilde{A}_1 + \widetilde{A}_2,
\end{equation}
where
\begin{equation*}
\widetilde{A}_1 = - 2 (\bm\lambda_1 - \widetilde{\bm\lambda}_1 )^\top \sumj \frac{\bm g(\bm X_{1,j},\bm\theta)}{ 1 + \breve{\bm\lambda}_1^\top \bm g(\bm X_{1,j},\bm\theta)}
\end{equation*}
with $\breve{\bm\lambda}_1 = \widetilde{a}_1 \bm\lambda_1 + (1-\widetilde{a}_1) \widetilde{\bm\lambda}_1$ for some $\widetilde{a}_1 \in (0,1)$, according to Taylor's expansion, and
\begin{equation*}
\widetilde{A}_2 = \eta_n \sum_{i'\in \mathcal{N}_1} \left( \| \bm\lambda_1 - \bm\lambda_{i'}\|_2 - \| \widetilde{\bm\lambda}_1 - \bm\lambda_{i'}\|_2 \right).
\end{equation*}
It is easy to see $\breve{\bm\lambda}_1 \in \mathcal{T}_{N^{-1}}^1 \setminus \widebar{\mathcal{T}}_{N^{-1}}$.
Similar to the analysis of $A_1$, it holds
\begin{equation}\label{P.Oracle.ii.A1}
|\widetilde{A}_1| = \| \bm\lambda_1 - \widetilde{\bm\lambda}_1 \|_2 o_p(\eta_n).
\end{equation}
From the definition of $\widetilde{\bm\lambda}_1$, it holds
\begin{equation}\label{P.Oracle.ii.A2}
\widetilde{A}_2 > \widetilde{c}_1 \eta_n \| \bm\lambda_1 - \widetilde{\bm\lambda}_1 \|_2
\end{equation}
for some constant $\widetilde{c}_1>0$.
Combining \eqref{P.Oracle.ii.split}, \eqref{P.Oracle.ii.A1} and \eqref{P.Oracle.ii.A2}, it holds
\begin{equation*}
\ell_{lasso}(\bm\Lambda) - \ell_{lasso}(\widetilde{\bm\Lambda}) = A_1 + A_2 \ge \leftsecond \eta_n + o_p(\eta_n) \rightsecond \| \bm\lambda_1 - \widetilde{\bm\lambda}_1 \|_2 > 0,
\end{equation*}
which implies
\begin{equation}\label{P.Orcal.ii}
\ell_{lasso}(\widetilde{\bm\Lambda}) < \ell_{lasso}(\bm\Lambda)
\end{equation}
with probability tending to $1$.
\eqref{P.Orcal.ii} together with the conclusion in (i) completes the proof of (ii).
That is, for any $\bm\Lambda \in \mathcal{D}_{N^{-1}} \setminus \widebar{\mathcal{D}}_{N^{-1}}$, it holds $\ell_{lasso}(\widehat{\bm\Lambda}^*) < \ell_{lasso}(\bm\Lambda)$ with probability tending to $1$.

Then, under the condition $\|\bm\theta - \bm\theta_0\|_2 = O(N^{-1/2})$ and $\eta_n/K\sqrt{n\log K} \to\infty$, we show $\Pr (\widehat{\bm\Lambda} = \widehat{\bm\Lambda}^*) \to 1$ as $n\to\infty$.
In this case, there exists a sequence $\varepsilon_n$ satisfying $\varepsilon_n \to 0$ and $\varepsilon_n^{-1}\eta_n/K\sqrt{n\log K} \to\infty$.
We complete the proof by analyzing (i') $\bm\Lambda \in \widebar{\mathcal{D}}_{\varepsilon_n}$ and (ii') $\bm\Lambda \in \mathcal{D}_{N^{-1}} \setminus \widebar{\mathcal{D}}_{\varepsilon_n}$.

(i') $\bm\Lambda \in \widebar{\mathcal{D}}_{\varepsilon_n}$:
Similar to the analysis above, we have $\ell_{lasso}(\bm\Lambda) - \ell_{lasso}(\widebar{\bm\Lambda}) = A_1 + A_2$ and $A_1 = -2n \sumi \bm G_i^\top (\bm\lambda_i - \widebar{\bm\lambda})$, which implies
\begin{equation}\label{P.Oracle.i'.GQ}
\Pr \left( \varepsilon_n \| \bm G_i \|_2 < \| \bm Q_i \|_2 \right) \to 1, ~\text{as}~n\to\infty,
\end{equation}
with $\bm Q_i = n^{-1} \sumj \bm g( \bm X_{i,j}, \bm\theta)$ being asymptotically normal distributed.
Under the condition $\|\bm\theta - \bm\theta_0\|_2 = O(N^{-1/2})$, it can be proved that $\max_{\coti} \|\bm Q_i\|_2 = O_p(n^{-1/2} \sqrt{\log K})$, which implies $\max_{\coti} \|\bm G_i\|_2 = O_p(\varepsilon_n^{-1} n^{-1/2} \sqrt{\log K})$ according to \eqref{P.Oracle.i'.GQ}.
Hence, it holds
\begin{equation*}
|A_1| \le nK \max_{\coti} \|\bm G_i\|_2 \max_{\coti} \| \bm\lambda_i - \widebar{\bm\lambda} \|_2 = \max_{\coti} \| \bm\lambda_i - \widebar{\bm\lambda} \|_2 O_p(\varepsilon_n^{-1} K \sqrt{n \log K})
\end{equation*}
for $\bm\Lambda \in \widebar{\mathcal{D}}_{\varepsilon_n}$.
From the condition $\eta_n/K\sqrt{n\log K} \to\infty$ and the choice of $\varepsilon_n$ that $\varepsilon_n \to 0$ and $\varepsilon_n^{-1}\eta_n/K\sqrt{n\log K} \to\infty$, we have
\begin{equation}\label{P.Oracle.i'.A1}
|A_1| = \max_{\coti} \| \bm\lambda_i - \widebar{\bm\lambda} \|_2 o_p(\eta_n)
\end{equation}
for $\bm\Lambda \in \widebar{\mathcal{D}}_{\varepsilon_n}$.
Similar to \eqref{P.Oracle.i.A2}, it can be proved that
\begin{equation}\label{P.Oracle.i'.A2}
A_2 \ge \eta_n \max_{\coti} \|\bm\lambda_i - \widebar{\bm\lambda} \|_2.
\end{equation}
Combining \eqref{P.Oracle.i'.A1} and \eqref{P.Oracle.i'.A2}, it holds $\ell_{lasso}(\bm\Lambda) - \ell_{lasso}(\widebar{\bm\Lambda}) >0$ for $\bm\Lambda \in \widebar{\mathcal{D}}_{\varepsilon_n}$ with $\bm\Lambda \neq \widebar{\bm\Lambda}$.
Also, we have $\ell_{lasso}(\widehat{\bm\Lambda}^*) < \ell_{lasso}(\widebar{\bm\Lambda})$ for $\forall \bm\Lambda \in \widebar{\mathcal{D}}_{\varepsilon_n}$ with $\widebar{\bm\Lambda} \neq \widehat{\bm\Lambda}^*$.
For these reasons, we have $\ell_{lasso}(\widehat{\bm\Lambda}^*) < \ell_{lasso}(\bm\Lambda)$ for $\forall \bm\Lambda \in \widebar{\mathcal{D}}_{\varepsilon_n}$ and $\bm\Lambda \neq \widehat{\bm\Lambda}^*$ with probability tending to $1$.

(ii') $\bm\Lambda \in \mathcal{D}_{N^{-1}} \setminus \widebar{\mathcal{D}}_{\varepsilon_n}$:
Given $\bm\Lambda \in \mathcal{D}_{N^{-1}} \setminus \widebar{\mathcal{D}}_{\varepsilon_n}$, define the index sets $\mathcal{I} := \{ (i,j):\, 1+ \bm\lambda_i^\top \bm g(\bm X_{i,j},\bm\theta) < \varepsilon_n \}$ and $\mathcal{I}^c := \{ (i,j):\, 1+ \bm\lambda_i^\top \bm g(\bm X_{i,j},\bm\theta) \ge \varepsilon_n \}$.
From the definition of $\ell_{lasso}(\bm\Lambda)$, it holds
\begin{equation}\label{P.Oracle.ii'.split}
\ell_{lasso}(\bm\Lambda) = \ell_{\mathcal{I}}(\bm\Lambda) + \ell_{lasso}'(\bm\Lambda),
\end{equation}
where
\begin{equation*}
\ell_{\mathcal{I}}(\bm\Lambda) = -2 \sum_{(i,j) \in \mathcal{I}} \log \leftsecond 1 + \bm\lambda_i^\top \bm g(\bm X_{i,j}, \bm\theta) \rightsecond
\end{equation*}
and
\begin{equation*}
\ell_{lasso}'(\bm\Lambda) = -2 \sum_{(i,j) \in \mathcal{I}^c} \log \leftsecond 1+ \bm\lambda_i^\top \bm g(\bm X_{i,j}, \bm\theta) \rightsecond + \sumi \sum_{\stackrel{i'\in \mathcal{N}_i}{i'>i}} \eta_n \| \bm\lambda_i - \bm\lambda_{i'}\|_2.
\end{equation*}
The definition of $\mathcal{I}$ implies
\begin{equation}\label{P.Oracle.ii'.1}
\ell_{\mathcal{I}}(\bm\Lambda) > -2 |\mathcal{I}| \log \varepsilon_n \to \infty.
\end{equation}
We denote the set $\mathcal{D}_{\varepsilon_n}$ which is generated according to the dataset $\{\bm X_{i,j}:\, (i,j) \in \mathcal{I}^c \}$ as $\mathcal{D}_{\varepsilon_n}'$.
It is shown that $\bm\Lambda \in \mathcal{D}_{\varepsilon_n}'$ according to the definition of $\mathcal{I}^c$.
From the statement in (i'), it holds
\begin{equation}\label{P.Oracle.ii'.2}
\ell_{lasso}'(\bm\Lambda) > \ell_{lasso}'(\widehat{\bm\Lambda}^{*'}),
\end{equation}
where $\widehat{\bm\Lambda}^{*'} = (\widehat{\bm\lambda}^{*'\top}, \ldots, \widehat{\bm\lambda}^{*'\top})^\top$ with
\begin{equation*}
\widehat{\bm\lambda}^{*'} = \arg\min_{\bm\lambda} -2 \sum_{(i,j) \in \mathcal{I}^c} \log \leftsecond 1 + \bm\lambda_i^\top \bm g(\bm X_{i,j}, \bm\theta) \rightsecond.
\end{equation*}

Following the same argument of \cite{owen1988empirical}, it can be shown that $\ell_{lasso}'(\widehat{\bm\lambda}^{*'}) = O_p(1)$ under the condition $\|\bm\theta - \bm\theta_0\|_2 = O(N^{-1/2})$.
This together with \eqref{P.Oracle.ii'.1} and \eqref{P.Oracle.ii'.2} implies that
\begin{equation}
\ell_{lasso}(\bm\Lambda) > \ell_{\mathcal{I}}(\bm\Lambda) + \ell_{lasso}'(\widehat{\bm\Lambda}^{*'}) \to \infty.
\end{equation}
Also, it holds $\ell_{lasso}(\widehat{\bm\lambda}^*) = O_p(1)$.
Therefore, we have $\ell_{lasso}(\widehat{\bm\Lambda}^*) < \ell_{lasso}(\bm\Lambda)$ for $\bm\Lambda \in \mathcal{D}_{N^{-1}} \setminus \widebar{\mathcal{D}}_{\varepsilon_n}$ with probability tending to $1$.

We complete the proof of Theorem 2.1.

\subsection{Proof of Theorem 2.2}

Theorem 2.2 holds based on Theorem 2.1 in this paper and Theorem 2 in \cite{qin1994empirical}.

\subsection{Proof of Lemma 3.1}

Lemma 3.1 can be easily proved according to Appendix A of \cite{hallac2015network}.

\subsection{Proof of Theorem 3.2}

Recalling the constrained objective function in PCM, we introduce the matrix representation for simplicity of notation.
Let $\mathbf{A}_{LR} = (\mathbf{A}_L^\top, \mathbf{A}_R^\top)^\top$, where $\mathbf{A}_L, \mathbf{A}_R \in \mathbb{R}^{M\times K}$ are matrices satisfying $(\mathbf{A}_{L})_{l \bullet} = \bm e_i^\top$ and $(\mathbf{A}_{R})_{l \bullet} = \bm e_{i'}^\top$, respectively, if $(\mathbf{A})_{l \bullet}$ is corresponding to the edge $e_{i,i'}$.
Let the matrix $\widetilde{\mathbf{A}}_{LR} \in \mathbb{R}^{2Mr \times Kr}$ satisfy $\widetilde{\mathbf{A}}_{LR} = \mathbf{A}_{LR} \otimes \mathbf{I}_r$, where $\otimes$ denotes the Kronecker product, and $\mathbf{I}_r \in \mathbb{R}^{r\times r}$ is the identity matrix.
We have the constraint $\bm C = \widetilde{\mathbf{A}}_{LR} \bm\Lambda \in \mathbb{R}^{2Mr}$.
Denote $\sumi \ell_i(\bm\lambda_i)$ by $\mathcal{L}(\bm\Lambda)$ and $\mathcal{P}_1 (\bm C) = \sum_{l=1}^M P_{\eta_n} (\widetilde{\bm\Delta}_l \bm C)$, where $\widetilde{\mathbf\Delta}_l = \bm\Delta_l \otimes \mathbf{I}_r$ and $\bm\Delta_l = (\mathbf{I}_M, -\mathbf{I}_M)_{l\bullet}$ for $l=1,\ldots,M$.
The optimization problem \eqref{eq:opt1} has the form
\begin{equation}\label{P.PCM.optimization}
\begin{aligned}
{}&{} \text{minimize} \quad \mathcal{L}(\bm\Lambda) + \mathcal{P}_1 (\bm C) \\
{}&{} \text{subject to} \quad \widetilde{\mathbf{A}}_{LR} \bm\Lambda - \bm C = \bm 0.
\end{aligned}
\end{equation}
By the augmented Lagrangian method, we have
\begin{equation*}
\mathcal{L}_{PCM} (\bm\Lambda,\bm C,\bm V) = \mathcal{L}(\bm\Lambda) + \mathcal{P}_1 (\bm C) + \bm V^\top (\widetilde{\mathbf{A}}_{LR} \bm\Lambda - \bm C) + \frac{\rho}{2} \|\widetilde{\mathbf{A}}_{LR} \bm\Lambda - \bm C\|_2^2,
\end{equation*}
where $\bm V \in \mathbb{R}^{2Mr}$ is the vector of dual variables corresponding to $\bm C$.

Next, we provide a convergence analysis for our proposed ADMM method for the problem.
Note that $\mathcal{L}(\bm\Lambda)$ and $\mathcal{P}_1 (\bm C)$ are convex functions, which ensures the convergence of the ADMM method for \eqref{P.PCM.optimization}.
Define
\begin{equation*}
\mathbf{H}_1:=
\begin{pmatrix}
\rho \widetilde{\mathbf{A}}_{LR}^\top \widetilde{\mathbf{A}}_{LR} & \mathbf{0} & \mathbf{0}\\
\mathbf{0} & \mathbf{0} & \mathbf{0}\\
\mathbf{0} & \mathbf{0} & \rho^{-1} \mathbf{I}_{2Mr}
\end{pmatrix},
\end{equation*}
and $\|\bm x\|_{\mathbf{H}_1} := \bm x^\top \mathbf{H}_1 \bm x$ for the symmetric and positive semi-definite matrix $\mathbf{H}_1$.
Let $\bm U_1^{(t)} = ( \bm\Lambda^{(t)\top},\bm C^{(t)\top}, \bm V^{(t)\top})^\top$ denote the $t$-th iteration generated by PCM, and $\bm U_{1*} = ( \bm\Lambda_{1*}^\top, \bm C_{*}^\top, \bm V_{*}^\top)^\top$ denote the KKT point for \eqref{P.PCM.optimization}.
From Theorems 6.1 in \cite{he2015non}, for any $t\ge 0$, we have
\begin{equation}\label{P.PCM.lemma.he}
\| \bm U_1^{(t)} - \bm U_1^{(t+1)} \|_{\mathbf{H}_1}^2 \le \frac{1}{t+1} \| \bm U_1^{(0)} - \bm U_{1*} \|_{\mathbf{H}_1}^2.
\end{equation}
Furthermore, by applying the results in Lemma 3.1 of \cite{he2015non}, it follows $\bm U_1^{(t+1)}$ is a solution of \eqref{P.PCM.optimization} if $\| \bm U_1^{(t)} - \bm U_1^{(t+1)} \|_{\mathbf{H}_1}^2 =0$.
Therefore, $\| \bm U_1^{(t)} - \bm U_1^{(t+1)} \|_{\mathbf{H}_1}^2$ can be viewed as an error measurement after $t$ iterations of PCM, and it is reasonable to conclude that the convergence rate of PCM is $O(1/t)$ according to \eqref{P.PCM.lemma.he}.
To be more specific, $\{ \bm U_1^{(t)} \}$ is bounded and there exists a converging subsequence $\{ \bm U_1^{(t_s)} \}$.
Let $\widebar{\bm U}_1 = \lim_{s\to\infty} \bm U_1^{(t_s)}$, and we next show $\widebar{\bm U}_1 = ( \widebar{\bm\Lambda}_1^\top, \widebar{\bm C}^\top, \widebar{\bm V}^\top)^\top$ is a KKT point.

From \eqref{P.PCM.lemma.he}, we have $\| \bm U_1^{(t)} - \bm U_1^{(t+1)} \|_{\mathbf{H}_1}^2 \to 0$, which implies $\bm V^{(t)} - \bm V^{(t+1)} \to \bm 0$ or equivalently,
\begin{equation}\label{P.PCM.residual}
\widetilde{\mathbf{A}}_{LR} \bm\Lambda^{(t+1)} - \bm C^{(t+1)} \to \bm 0, ~\text{as}~ t\to\infty.
\end{equation}
Consequently, $\| \bm U_1^{(t)} - \bm U_{1*} \|_{\mathbf{H}_1}^2$ also converges to a constant.
By passing limit on \eqref{P.PCM.residual} over the subsequence, we have
\begin{equation}\label{P.PCM.KKT.1}
\widetilde{\mathbf{A}}_{LR} \widebar{\bm\Lambda}_1 - \widebar{\bm C} = \bm 0.
\end{equation}
Furthermore, from the generation mechanism of $\bm U^{(t+1)}$, we have
\begin{equation}\label{P.PCM.lim}
\begin{aligned}
\bm 0 {}&{} \in \partial \mathcal{P}_1(\bm C^{(t+1)}) - \bm V^{(t+1)},\\
\bm 0 {}&{} = \nabla \mathcal{L}(\bm\Lambda^{(t+1)}) + \widetilde{\mathbf{A}}_{LR}^\top \bm V^{(t+1)} - \rho \widetilde{\mathbf{A}}_{LR}^\top (\bm C^{(t)} - \bm C^{(t+1)}),
\end{aligned}
\end{equation}
where $\partial \mathcal{P}_1(\bm C^{(t+1)})$ denotes the subdifferential of the function $\mathcal{P}_1$ at the point $\bm C^{(t+1)}$.
\eqref{P.PCM.lemma.he} implies $\| \bm U_1^{(t)} - \bm U_1^{(t+1)} \|_{\mathbf{H}_1}^2 \to 0$, and hence we have $\bm C^{(t)} - \bm C^{(t+1)} \to \bm 0$.
Taking limit on \eqref{P.PCM.lim} over the subsequence, we have
\begin{equation}\label{P.PCM.KKT.2}
\begin{aligned}
\bm 0 {}&{} \in \partial \mathcal{P}_1(\widebar{\bm C}) - \widebar{\bm V},\\
\bm 0 {}&{} = \nabla \mathcal{L}(\widebar{\bm\Lambda}_1) + \widetilde{\mathbf{A}}_{LR}^\top \widebar{\bm V}.
\end{aligned}
\end{equation}
By \eqref{P.PCM.KKT.1} and \eqref{P.PCM.KKT.2}, it follows that $\widebar{\bm U}_1$ is a KKT point, and we can take $\bm U_{1*} = \widebar{\bm U}_1$.
From $\bm U_1^{(t_s)} \to \widebar{\bm U}_1$ in $s$ and the convergence of $\| \bm U_1^{(t)} - \bm U_{1*} \|_{\mathbf{H}_1}^2$, it follows $\| \bm U_1^{(t)} - \bm U_{1*} \|_{\mathbf{H}_1}^2 \to 0$ in $t$.

\subsection{Proof of Corollary 3.3}
From the proof of Theorem 2.1, we know $\bm\Lambda = \widehat{\bm\Lambda}^*$ is the unique global minimizer of the constrained optimization problem \eqref{P.PCM.optimization} satisfying $\bm C = \widetilde{\mathbf{A}}_{LR} \bm\Lambda$.
This together with the KKT conditions proves $\bm\Lambda_{1*} = \widehat{\bm\Lambda}^*$ and $\bm\lambda_i^{(t)} \to \widehat{\bm\lambda}^*$ uniformly for $\coti$.
This together with Theorem 2.2 completes the proof.

\subsection{Proof of Theorem 3.4}

Considering that the objective function in MAOM is changed in each iteration, the convergence property of MAOM need to be deduced carefully.
To our best knowledge, literature about convergence analysis of ADMM always focus on fixed objective function (e.g. \cite{deng2016global}), while the convergence property of modified approximated objective function in each iteration has not been proposed.
In the following, we provide the convergence property for MAOM.

Recalling the constrained objective function in MAOM, for notation simplicity, we denote $\mathcal{P}_2 (\bm Z) = \sum_{l=1}^M P_\eta (\widetilde{\mathbf{O}}_l \bm Z)$.
The constrained optimization problem \eqref{eq:opt2} has the form
\begin{equation}\label{P.MAOM.optimization}
\begin{aligned}
{}&{} \text{minimize} \quad \mathcal{L}(\bm\Lambda) + \mathcal{P}_2 (\bm Z)\\
{}&{} \text{subject to} \quad \widetilde{\mathbf{A}} \bm\Lambda - \bm Z = \bm 0.
\end{aligned}
\end{equation}
The corresponding augmented Lagrange problem is $\min_{\bm\Lambda, \bm Z, \bm T} \mathcal{L}_2 (\bm\Lambda, \bm Z, \bm T)$, where
\begin{equation*}
\mathcal{L}_2 (\bm\Lambda, \bm Z, \bm T) = \mathcal{L}(\bm\Lambda) + \mathcal{P}_2 (\bm Z) + \bm T^\top (\widetilde{\mathbf{A}} \bm\Lambda - \bm Z) + \frac{\rho}{2} \|\widetilde{\mathbf{A}} \bm\Lambda - \bm Z\|_2^2.
\end{equation*}
By MAOM, the update strategy for parameters is
\begin{equation}\label{P.MAOM.algorithm}
\begin{aligned}
\bm Z^\iterN = {}&{} \arg\min_{\bm Z} \widehat{\mathcal{L}}_2^\iterO (\bm\Lambda^\iterO, \bm Z, \bm T^\iterO),\\
\bm\Lambda^\iterN = {}&{} \arg\min_{\bm\Lambda} \widehat{\mathcal{L}}_2^\iterO (\bm\Lambda, \bm Z^\iterN, \bm T^\iterO),\\
\bm T^\iterN = {}&{} \bm T^\iterO + \rho (\widetilde{\mathbf{A}} \bm\Lambda^\iterN - \bm Z^\iterN),
\end{aligned}
\end{equation}
where
\begin{equation*}
\widehat{\mathcal{L}}_2^\iterO (\bm\Lambda, \bm Z, \bm T) = \widehat{\mathcal{L}}^{(t)} (\bm\Lambda;\widetilde{\mathbf{Q}}) + \mathcal{P}_2 (\bm Z) + \bm T^\top (\widetilde{\mathbf{A}} \bm\Lambda - \bm Z) + \frac{\rho}{2} \|\widetilde{\mathbf{A}} \bm\Lambda - \bm Z\|_2^2,
\end{equation*}
and
\begin{equation*}
\begin{aligned}
\widehat{\mathcal{L}}^{(t)} (\bm\Lambda;\widetilde{\mathbf{Q}}) ={}&{} \mathcal{L} (\bm\Lambda^{(t)}) + (\bm\Lambda - \bm\Lambda^{(t)})^\top \nabla \mathcal{L} (\bm\Lambda^{(t)})\\
{}&{} + \frac{1}{2} (\bm\Lambda - \bm\Lambda^{(t)})^\top \nabla^2 \mathcal{L} (\bm\Lambda^{(t)}) (\bm\Lambda - \bm\Lambda^{(t)}) + \frac{1}{2} (\bm\Lambda - \bm\Lambda^{(t)})^\top \widetilde{\mathbf{Q}} (\bm\Lambda - \bm\Lambda^{(t)}).
\end{aligned}
\end{equation*}
Define
\begin{equation*}
\mathbf{H}_2:=
\begin{pmatrix}
\widetilde{\mathbf{Q}} + \rho \widetilde{\mathbf{A}}^\top \widetilde{\mathbf{A}} & \mathbf{0} & \mathbf{0}\\
\mathbf{0} & \mathbf{0} & \mathbf{0}\\
\mathbf{0} & \mathbf{0} & \rho^{-1} \mathbf{I}_{Mr}
\end{pmatrix},
\end{equation*}
and $\|\bm x\|_{\mathbf{H}_2} := \bm x^\top \mathbf{H}_2 \bm x$ for the symmetric and positive semi-definite matrix $\mathbf{H}_2$.
Let $\bm U_2^{(t)} = ( \bm\Lambda^{\iterO\top},\bm Z^{\iterO\top}, \bm T^{\iterO\top})^\top$ denote the $t$-th iteration generated by MAOM, and $\bm U_{2*} = ( \bm\Lambda_{2*}^\top, \bm Z_{*}^\top, \bm T_{*}^\top)^\top$ denote the KKT point for \eqref{P.MAOM.optimization}.
We next prove $\| \bm U_2^{(t)} - \bm U_{2*} \|_{\mathbf{H}_2}^2 \to 0$ in $t$.

From the update strategy \eqref{P.MAOM.algorithm}, we have
\begin{equation}\label{P.MAOM.diffL}
\begin{aligned}
\bm 0 {}&{} \in \partial \mathcal{P}_2 (\bm Z^\iterN) - \bm T^\iterO - \rho ( \widetilde{\mathbf{A}} \bm\Lambda^\iterO - \bm Z^\iterN),\\
\bm 0 {}&{} = \nabla \widehat{\mathcal{L}}^\iterO (\bm\Lambda^\iterN;\widetilde{\mathbf{Q}}) + \widetilde{\mathbf{A}}^\top \bm T^\iterO + \rho \widetilde{\mathbf{A}}^\top ( \widetilde{\mathbf{A}} \bm\Lambda^\iterN - \bm Z^\iterN),
\end{aligned}
\end{equation}
where
\begin{equation}\label{P.MAOM.diffHatL.1}
\nabla \widehat{\mathcal{L}}^\iterO (\bm\Lambda^\iterN;\widetilde{\mathbf{Q}}) = \nabla \mathcal{L} (\bm\Lambda^\iterO) + \nabla^2 \mathcal{L} (\bm\Lambda^\iterO) ( \bm\Lambda^\iterN - \bm\Lambda^\iterO) + \widetilde{\mathbf{Q}} ( \bm\Lambda^\iterN - \bm\Lambda^\iterO).
\end{equation}
From the definition of the KKT point $\bm U_{2*} = ( \bm\Lambda_{2*}^\top, \bm Z_{*}^\top, \bm T_{*}^\top)^\top$, we have
\begin{equation}\label{P.MAOM.KKTpoint}
\begin{aligned}
\bm 0 {}&{} \in \partial \mathcal{P}_2 (\bm Z_*) - \bm T_*,\\
\bm 0 {}&{} = \nabla \mathcal{L} (\bm\Lambda_{2*}) + \widetilde{\mathbf{A}}^\top \bm T_*,\\
\bm 0 {}&{} = \widetilde{\mathbf{A}} \bm\Lambda_{2*} - \bm Z_*.
\end{aligned}
\end{equation}
From the convexity of $\mathcal{P}_2 (\bm Z)$ and the strong convexity of $\mathcal{L} (\bm\Lambda)$, we have
\begin{equation}\label{P.MAOM.convexity}
\begin{aligned}
(\bm Z_1 - \bm Z_2)^\top \leftsecond \partial \mathcal{P}_2 (\bm Z_1) - \partial \mathcal{P}_2 (\bm Z_1) \rightsecond \ge v_{\mathcal{P}} \| \bm Z_1 - \bm Z_2 \|_2^2, ~\forall \bm Z_1,\bm Z_2,\\
(\bm\Lambda_1 - \bm\Lambda_2)^\top \leftsecond \nabla \mathcal{L} (\bm\Lambda_1) - \nabla \mathcal{L} (\bm\Lambda_1) \rightsecond \ge v_{\mathcal{L}} \| \bm\Lambda_1 - \bm\Lambda_2 \|_2^2, ~\forall \bm\Lambda_1,\bm\Lambda_2,
\end{aligned}
\end{equation}
where $v_{\mathcal{P}} \ge 0$ and $v_{\mathcal{L}} > 0$.
By \eqref{P.MAOM.diffL}, \eqref{P.MAOM.KKTpoint} and \eqref{P.MAOM.convexity}, we have
\begin{equation}\label{P.MAOM.Important}
\begin{aligned}
(\bm Z^\iterN - \bm Z_*)^\top \leftsecond \bm T^\iterO + \rho ( \widetilde{\mathbf{A}} \bm\Lambda^\iterO - \bm Z^\iterN) - \bm T_* \rightsecond \ge {}&{} v_{\mathcal{P}} \| \bm Z^\iterN - \bm Z_* \|_2^2,\\
(\bm\Lambda^\iterN - \bm\Lambda_{2*})^\top \leftsecond \nabla \mathcal{L} (\bm\Lambda^\iterN) + \widetilde{\mathbf{A}}^\top \bm T_* \rightsecond \ge {}&{} v_{\mathcal{L}} \| \bm\Lambda^\iterN - \bm\Lambda_{2*} \|_2^2,
\end{aligned}
\end{equation}
where 
\begin{equation}\label{P.MAOM.diffHatL.2}
\begin{aligned}
\nabla \mathcal{L} (\bm\Lambda^\iterN) = {}&{} \leftsecond \nabla \mathcal{L} (\bm\Lambda^\iterN) - \nabla \mathcal{L} (\bm\Lambda^\iterO) \rightsecond - \leftsecond \nabla^2 \mathcal{L} (\bm\Lambda^\iterO) + \widetilde{\mathbf{Q}} \rightsecond ( \bm\Lambda^\iterN - \bm\Lambda^\iterO)\\
{}&{} - \widetilde{\mathbf{A}}^\top \bm T^\iterO - \rho \widetilde{\mathbf{A}}^\top ( \widetilde{\mathbf{A}} \bm\Lambda^\iterN - \bm Z^\iterN)
\end{aligned}
\end{equation}
by \eqref{P.MAOM.diffHatL.1}.
By adding the two inequations in \eqref{P.MAOM.Important} together, it follows
\begin{equation}\label{P.MAOM.add.1}
\begin{aligned}
{}&{} (\bm Z^\iterN - \bm Z_*)^\top \leftsecond \bm T^\iterO + \rho ( \widetilde{\mathbf{A}} \bm\Lambda^\iterO - \bm Z^\iterN) - \bm T_* \rightsecond\\
{}&{} + (\bm\Lambda^\iterN - \bm\Lambda_{2*})^\top \leftthird \leftsecond \nabla \mathcal{L} (\bm\Lambda^\iterN) - \nabla \mathcal{L} (\bm\Lambda^\iterO) \rightsecond \rightdot\\
{}&{} - \leftdot \leftsecond \nabla^2 \mathcal{L} (\bm\Lambda^\iterO) + \widetilde{\mathbf{Q}} \rightsecond ( \bm\Lambda^\iterN - \bm\Lambda^\iterO) \rightthird\\
{}&{} - (\bm\Lambda^\iterN - \bm\Lambda_{2*})^\top \leftsecond \widetilde{\mathbf{A}}^\top \bm T^\iterO + \rho \widetilde{\mathbf{A}}^\top ( \widetilde{\mathbf{A}} \bm\Lambda^\iterN - \bm Z^\iterN) - \widetilde{\mathbf{A}}^\top \bm T_* \rightsecond\\
\ge {}&{} v_{\mathcal{P}} \| \bm Z^\iterN - \bm Z_* \|_2^2 + v_{\mathcal{L}} \| \bm\Lambda^\iterN - \bm\Lambda_{2*} \|_2^2.
\end{aligned}
\end{equation}
From \eqref{P.MAOM.algorithm}, we have
\begin{equation*}
\bm T^\iterN = \bm T^\iterO + \rho (\widetilde{\mathbf{A}} \bm\Lambda^\iterN - \bm Z^\iterN).
\end{equation*}
Hence, \eqref{P.MAOM.add.1} can be simplified to
\begin{equation}\label{P.MAOM.Xi}
\Xi_1 + \Xi_2 + \Xi_3 + \Xi_4 \ge v_{\mathcal{P}} \| \bm Z^\iterN - \bm Z_* \|_2^2 + v_{\mathcal{L}} \| \bm\Lambda^\iterN - \bm\Lambda_{2*} \|_2^2,
\end{equation}
where
\begin{equation}\label{P.MAOM.XiDetail}
\begin{aligned}
\Xi_1 {}&{} = (\bm Z^\iterN - \bm Z_*)^\top (\bm T^\iterN - \bm T_*) - (\bm\Lambda^\iterN - \bm\Lambda_{2*})^\top \widetilde{\mathbf{A}}^\top (\bm T^\iterN - \bm T_*),\\
\Xi_2 {}&{} = \rho (\bm Z^\iterN - \bm Z_*)^\top \widetilde{\mathbf{A}} (\bm\Lambda^\iterO - \bm\Lambda^\iterN),\\
\Xi_3 {}&{} = (\bm\Lambda^\iterN - \bm\Lambda_{2*})^\top \widetilde{\mathbf{Q}} ( \bm\Lambda^\iterO - \bm\Lambda^\iterN),\\
\Xi_4 {}&{} = (\bm\Lambda^\iterN - \bm\Lambda_{2*})^\top \leftthird \leftsecond \nabla \mathcal{L} (\bm\Lambda^\iterN) - \nabla \mathcal{L} (\bm\Lambda^\iterO) \rightsecond + \nabla^2 \mathcal{L} (\bm\Lambda^\iterO) ( \bm\Lambda^\iterO - \bm\Lambda^\iterN) \rightthird.
\end{aligned}
\end{equation}
Combining \eqref{P.MAOM.algorithm} and \eqref{P.MAOM.KKTpoint}, we get
\begin{equation}\label{P.MAOM.link.1}
\frac{1}{\rho} (\bm T^\iterO - \bm T^\iterN) = (\bm Z^\iterN - \bm Z_*) - \widetilde{\mathbf{A}} (\bm\Lambda^\iterN - \bm\Lambda_{2*}).
\end{equation}
This proves
\begin{equation}\label{P.MAOM.Xi1}
\Xi_1 = \frac{1}{\rho} (\bm T^\iterN - \bm T_*)^\top (\bm T^\iterO - \bm T^\iterN),
\end{equation}
and
\begin{equation}\label{P.MAOM.Xi2}
\Xi_2 = (\bm T^\iterO - \bm T^\iterN)^\top \widetilde{\mathbf{A}} (\bm\Lambda^\iterO - \bm\Lambda^\iterN) + \rho (\bm\Lambda^\iterN - \bm\Lambda_{2*})^\top \widetilde{\mathbf{A}}^\top \widetilde{\mathbf{A}} (\bm\Lambda^\iterO - \bm\Lambda^\iterN).
\end{equation}
For $\Xi_4$, by the basic inequality, we have
\begin{equation*}
2 |\Xi_4| \le \frac{1}{w_1} \| \bm\Lambda^\iterN - \bm\Lambda_{2*} \|_2^2 + w_1 \| \bm \xi(\bm\Lambda^\iterO,\bm\Lambda^\iterN) \|_2^2,~\text{for}~w_1>0,
\end{equation*}
where
\begin{equation*}
\bm \xi(\bm\Lambda^\iterO,\bm\Lambda^\iterN) = \leftsecond \nabla \mathcal{L} (\bm\Lambda^\iterN) - \nabla \mathcal{L} (\bm\Lambda^\iterO) \rightsecond + \nabla^2 \mathcal{L} (\bm\Lambda^\iterO) ( \bm\Lambda^\iterO - \bm\Lambda^\iterN).
\end{equation*}
From Taylor's expansion, we have
\begin{equation*}
\bm \xi(\bm\Lambda^\iterO,\bm\Lambda^\iterN) = \frac{1}{2} \nabla^3 \mathcal{L}(a\bm\Lambda^\iterO + (1-a)\bm\Lambda^\iterN) \left( \bm\Lambda^\iterN - \bm\Lambda^\iterO \right)^{\otimes 2}.
\end{equation*}
for some $a\in(0,1)$.
From $\| \nabla^3 \mathcal{L}(\bm\Lambda) \|_2 = O_p(1)$, we have
\begin{equation*}
\| \bm \xi(\bm\Lambda^\iterO,\bm\Lambda^\iterN) \|_2^2 = O_p( \| \bm\Lambda^\iterO - \bm\Lambda^\iterN \|_2^4).
\end{equation*}
Since the initial point is $\bm\Lambda_0 = \bm 0$ and the update procedure ensures that $\bm\Lambda^{(t)}$ converges to $\widehat{\bm\Lambda}^* = O_p(N^{-1/2})$ as $t\to\infty$, it holds that $\| \bm\Lambda^\iterO - \bm\Lambda^\iterN \|_2^2 = o_p(1)$, and $w_1 \| \bm \xi(\bm\Lambda^\iterO,\bm\Lambda^\iterN) \|_2^2 = o_p( \| \bm\Lambda^\iterO - \bm\Lambda^\iterN \|_2^2)$.
This proves
\begin{equation}\label{P.MAOM.xi4}
|\Xi_4| \le \frac{1}{2w_1} \| \bm\Lambda^\iterN - \bm\Lambda_{2*} \|_2^2 + o_p( \| \bm\Lambda^\iterO - \bm\Lambda^\iterN \|_2^2),~\text{for}~w_1>0,
\end{equation}
Combining \eqref{P.MAOM.Xi}, \eqref{P.MAOM.XiDetail}, \eqref{P.MAOM.Xi1}, \eqref{P.MAOM.Xi2}, and \eqref{P.MAOM.xi4}, we obtain
\begin{equation*}
\begin{aligned}
{}&{} (\bm\Lambda^\iterN - \bm\Lambda_{2*})^\top (\widetilde{\mathbf{Q}} + \rho \widetilde{\mathbf{A}}^\top \widetilde{\mathbf{A}}) (\bm\Lambda^\iterO - \bm\Lambda^\iterN) + \frac{1}{\rho} (\bm T^\iterN - \bm T_*)^\top (\bm T^\iterO - \bm T^\iterN)\\
{}&{} + (\bm T^\iterO - \bm T^\iterN)^\top \widetilde{\mathbf{A}} (\bm\Lambda^\iterO - \bm\Lambda^\iterN) + \frac{1}{2w_1} \| \bm\Lambda^\iterN - \bm\Lambda_{2*} \|_2^2 + o_p( \| \bm\Lambda^\iterO - \bm\Lambda^\iterN \|_2^2)\\
\ge {}&{} v_{\mathcal{P}} \| \bm Z^\iterN - \bm Z_* \|_2^2 + v_{\mathcal{L}} \| \bm\Lambda^\iterN - \bm\Lambda_{2*} \|_2^2,
\end{aligned}
\end{equation*}
which implies
\begin{equation}\label{P.MAOM.add.3}
\begin{aligned}
{}&{} (\bm U^\iterN - \bm U_{2*})^\top \mathbf{H}_2 (\bm U^\iterO - \bm U^\iterN)\\
\ge {}&{} v_{\mathcal{P}} \| \bm Z^\iterN - \bm Z_* \|_2^2 + v_{\mathcal{L}} \| \bm\Lambda^\iterN - \bm\Lambda_{2*} \|_2^2 - \frac{1}{2w_1} \| \bm\Lambda^\iterN - \bm\Lambda_{2*} \|_2^2\\
{}&{} - (\bm T^\iterO - \bm T^\iterN)^\top \widetilde{\mathbf{A}} (\bm\Lambda^\iterO - \bm\Lambda^\iterN) + o_p( \| \bm\Lambda^\iterO - \bm\Lambda^\iterN \|_2^2).
\end{aligned}
\end{equation}
By $\|\bm a - \bm c\|_{\mathbf{H}_2}^2 - \|\bm b - \bm c\|_{\mathbf{H}_2}^2 = 2(\bm a - \bm c)^\top \mathbf{H}_2 (\bm a - \bm b) - \|\bm a - \bm b\|_{\mathbf{H}_2}^2$, we have
\begin{equation}\label{P.MAOM.link.2}
\begin{aligned}
{}&{} 2(\bm U^\iterN - \bm U_{2*})^\top \mathbf{H}_2 (\bm U^\iterO - \bm U^\iterN)\\
= {}&{} \|\bm U^\iterO - \bm U_{2*}\|_{\mathbf{H}_2}^2 -
\|\bm U^\iterN - \bm U_{2*}\|_{\mathbf{H}_2}^2 - \|\bm U^\iterO - \bm U^\iterN\|_{\mathbf{H}_2}^2.
\end{aligned}
\end{equation}
Combining \eqref{P.MAOM.add.3} and \eqref{P.MAOM.link.2}, we have
\begin{equation}\label{P.MAOM.add.4}
\begin{aligned}
{}&{} \|\bm U^\iterO - \bm U_{2*}\|_{\mathbf{H}_2}^2 -
\|\bm U^\iterN - \bm U_{2*}\|_{\mathbf{H}_2}^2\\
\ge {}&{} \|\bm U^\iterO - \bm U^\iterN\|_{\mathbf{H}_2}^2 + 2 v_{\mathcal{P}} \| \bm Z^\iterN - \bm Z_* \|_2^2\\
{}&{} + 2 v_{\mathcal{L}} \| \bm\Lambda^\iterN - \bm\Lambda_{2*} \|_2^2 - \frac{1}{w_1} \| \bm\Lambda^\iterN - \bm\Lambda_{2*} \|_2^2\\
{}&{} - 2 (\bm T^\iterO - \bm T^\iterN)^\top \widetilde{\mathbf{A}} (\bm\Lambda^\iterO - \bm\Lambda^\iterN) + o_p( \| \bm\Lambda^\iterO - \bm\Lambda^\iterN \|_2^2).
\end{aligned}
\end{equation}
By the basic inequality, we have
\begin{equation}\label{P.MAOM.link.3}
\begin{aligned}
{}&{} -2 (\bm T^\iterO - \bm T^\iterN)^\top \widetilde{\mathbf{A}} (\bm\Lambda^\iterO - \bm\Lambda^\iterN)\\
\ge {}&{} - w_2 \| \widetilde{\mathbf{A}} (\bm\Lambda^\iterO - \bm\Lambda^\iterN) \|_2^2 - \frac{1}{w_2} \| \bm T^\iterO - \bm T^\iterN\|_2^2, ~\text{for}~ w_2>0.
\end{aligned}
\end{equation}
Recalling the definition of $\mathbf{H}_2$, if we choose $w_2 > \rho$ satisfying $\widetilde{\mathbf{Q}} +(\rho-w_2) \widetilde{\mathbf{A}}^\top \widetilde{\mathbf{A}} \succ \mathbf{0}$, it holds
\begin{equation}\label{P.MAOM.add.5.1}
\begin{aligned}
{}&{} \|\bm U^\iterO - \bm U^\iterN\|_{\mathbf{H}_2}^2 - w_2 \| \widetilde{\mathbf{A}} (\bm\Lambda^\iterO - \bm\Lambda^\iterN) \|_2^2\\
{}&{} - \frac{1}{w_2} \| \bm T^\iterO - \bm T^\iterN\|_2^2 + o_p( \| \bm\Lambda^\iterO - \bm\Lambda^\iterN \|_2^2)\\
\ge {}&{} c_1 \|\bm U^\iterO - \bm U^\iterN\|_{\mathbf{H}_2}^2,
\end{aligned}
\end{equation}
for some $c_1>0$.
Under the definition of $\widetilde{\mathbf{Q}}$, it holds $\widetilde{\mathbf{Q}}\succ \mathbf{0}$, and there exists a $w_2$ satisfying the above inequation.
Additionally, if we choose $w_1>1/ 2 v_{\mathcal{L}}$, it holds
\begin{equation}\label{P.MAOM.add.5.2}
2 v_{\mathcal{L}} \| \bm\Lambda^\iterN - \bm\Lambda_{2*} \|_2^2 - \frac{1}{w_1} \| \bm\Lambda^\iterN - \bm\Lambda_{2*} \|_2^2 = c_2 \| \bm\Lambda^\iterN - \bm\Lambda_{2*} \|_2^2,
\end{equation}
for some $c_2 = 2v_{\mathcal{L}} - 1/w_1 >0$.
Combining \eqref{P.MAOM.add.4}, \eqref{P.MAOM.link.3}, \eqref{P.MAOM.add.5.1}, and \eqref{P.MAOM.add.5.2}, it follows
\begin{equation}\label{P.MAOM.conclusion}
\begin{aligned}
{}&{} \|\bm U^\iterO - \bm U_{2*}\|_{\mathbf{H}_2}^2 -
\|\bm U^\iterN - \bm U_{2*}\|_{\mathbf{H}_2}^2\\
\ge {}&{} c_1 \|\bm U^\iterO - \bm U^\iterN\|_{\mathbf{H}_2}^2 + 2 v_{\mathcal{P}} \| \bm Z^\iterN - \bm Z_* \|_2^2 + c_2 \| \bm\Lambda^\iterN - \bm\Lambda_{2*} \|_2^2,
\end{aligned}
\end{equation}
where $c_1$ and $c_2$ are positive constants.

From \eqref{P.MAOM.conclusion}, we conclude that $\|\bm U^\iterO - \bm U_{2*}\|_{\mathbf{H}_2}^2$ is monotonically nonincreasing and thus converging, and due to $c_1>0$, we have $\|\bm U^\iterO - \bm U^\iterN\|_{\mathbf{H}_2}^2 \to 0$.
Similar to the operation in the proof of Theorem 3.1, it is easy to prove $\| \bm U_2^\iterO - \bm U_{2*} \|_{\mathbf{H}_2}^2 \to 0$, as $t \to \infty$.

\subsection{Proof of Corollary 3.5}
Similar to the proof of Corollary 3.2, we have $\bm\Lambda_{2*} = \widehat{\bm\Lambda}^*$ and $\bm\lambda_i^\iterO \to \widehat{\bm\lambda}^*$ as $t \to \infty$ uniformly for $\coti$.
This together with Theorem 2.2 completes the proof.

\subsection{Proof of Theorem 3.6}

(a) From \cite{zhang2001largest}, it follows that the $(K-1)$-th largest eigenvalue of $\mathbf{A}_{T}^\top \mathbf{A}_{T}$ is positive.
Hence, the incidence matrix is of full row rank.

(b) We complete the proof following the framework of \cite{deng2016global}.
According to \eqref{P.MAOM.conclusion}, to prove $\|\bm U^\iterO - \bm U_{2*}\|_{\mathbf{H}_2}^2 \ge (1+c_3) \|\bm U^\iterN - \bm U_{2*}\|_{\mathbf{H}_2}^2$ for some $c_3>0$, it is sufficient to show
\begin{equation}\label{P.Linear.goal}
\begin{aligned}
{}&{} c_1 \|\bm U^\iterO - \bm U^\iterN\|_{\mathbf{H}_2}^2 + 2 v_{\mathcal{P}} \| \bm Z^\iterN - \bm Z_* \|_2^2 + c_2 \| \bm\Lambda^\iterN - \bm\Lambda_{2*} \|_2^2\\
\ge {}&{} c_3 \|\bm U^\iterN - \bm U_{2*}\|_{\mathbf{H}_2}^2.
\end{aligned}
\end{equation}
From \eqref{P.MAOM.diffHatL.2} and \eqref{P.MAOM.KKTpoint}, together with Lipschitz continuity of $\nabla \mathcal{L}$, we have
\begin{equation}\label{P.Linear.basis}
\begin{aligned}
{}&{} \| \nabla \mathcal{L} (\bm\Lambda^\iterO) - \nabla \mathcal{L} (\bm\Lambda^\iterN) + \leftsecond \nabla^2 \mathcal{L} (\bm\Lambda^\iterO) + \widetilde{\mathbf{Q}} \rightsecond ( \bm\Lambda^\iterN - \bm\Lambda^\iterO)\\
{}&{} + \widetilde{\mathbf{A}}^\top ( \bm T^\iterN - \bm T_*) \|_2^2\\
= {}&{} \| \nabla \mathcal{L} (\bm\Lambda_{2*}) - \nabla \mathcal{L} (\bm\Lambda^\iterN) \|_2^2\\
\le {}&{} L_{\mathcal{L}}^2 \| \bm\Lambda^\iterN - \bm\Lambda_{2*} \|_2^2.
\end{aligned}
\end{equation}
By the inequality
\begin{equation*}
\| \bm a + \bm b\|_2^2 \ge (1-\frac{1}{w_3}) \|\bm a\|_2^2 + (1-w_3) \|\bm b\|_2^2,~ \text{for}~ w_3>0,
\end{equation*}
we get
\begin{equation}\label{P.Linear.inequation}
\begin{aligned}
{}&{} \| \nabla \mathcal{L} (\bm\Lambda^\iterO) - \nabla \mathcal{L} (\bm\Lambda^\iterN) + \leftsecond \nabla^2 \mathcal{L} (\bm\Lambda^\iterO) + \widetilde{\mathbf{Q}} \rightsecond ( \bm\Lambda^\iterN - \bm\Lambda^\iterO)\\
{}&{}+ \widetilde{\mathbf{A}}^\top ( \bm T^\iterN - \bm T_*) \|_2^2\\
\ge {}&{} (1- \frac{1}{w_3}) \| \nabla \mathcal{L} (\bm\Lambda^\iterO) - \nabla \mathcal{L} (\bm\Lambda^\iterN) + \leftsecond \nabla^2 \mathcal{L} (\bm\Lambda^\iterO) + \widetilde{\mathbf{Q}} \rightsecond ( \bm\Lambda^\iterN - \bm\Lambda^\iterO) \|_2^2\\
{}&{} + (1-w_3) \| \widetilde{\mathbf{A}}^\top ( \bm T^\iterN - \bm T_*) \|_2^2\\
\ge {}&{} (1- \frac{1}{w_3})(1- w_4) \| \widetilde{\mathbf{Q}} ( \bm\Lambda^\iterN - \bm\Lambda^\iterO) \|_2^2 + (1- \frac{1}{w_3})(1- \frac{1}{w_4}) \| \bm \xi(\bm\Lambda^\iterO,\bm\Lambda^\iterN) \|_2^2\\
{}&{} + (1-w_3) \| \widetilde{\mathbf{A}}^\top ( \bm T^\iterN - \bm T_*) \|_2^2
\end{aligned}
\end{equation}
for some $w_3,w_4>0$.
In order to ensure that $1-w_3, 1-w_4>0$, we take $0< w_3, w_4 <1$.
By \eqref{P.Linear.basis} and \eqref{P.Linear.inequation}, we have
\begin{equation*}
\begin{aligned}
{}&{} \| \widetilde{\mathbf{A}}^\top ( \bm T^\iterN - \bm T_*) \|_2^2\\
\le {}&{} - \frac{1}{1-w_3} (1- \frac{1}{w_3})(1- w_4) \| \widetilde{\mathbf{Q}} ( \bm\Lambda^\iterN - \bm\Lambda^\iterO) \|_2^2 \\
{}&{} + \frac{1}{1-w_3} L_{\mathcal{L}}^2 \| \bm\Lambda^\iterN - \bm\Lambda_{2*} \|_2^2 - \frac{1}{1-w_3} (1- \frac{1}{w_3})(1- \frac{1}{w_4}) \| \bm \xi(\bm\Lambda^\iterO,\bm\Lambda^\iterN) \|_2^2\\
\le {}&{} \frac{1}{w_3} (1- w_4) \| \widetilde{\mathbf{Q}} \|_2^2 \| \bm\Lambda^\iterN - \bm\Lambda^\iterO \|_2^2 + \frac{1}{1-w_3} L_{\mathcal{L}}^2 \| \bm\Lambda^\iterN - \bm\Lambda_{2*} \|_2^2.
\end{aligned}
\end{equation*}
Since $\widetilde{\mathbf{A}}$ has full row rank, it follows
\begin{equation}\label{P.Linear.expansion.1}
\| \bm T^\iterN - \bm T_* \|_2^2 \le \frac{(1- w_4) \| \widetilde{\mathbf{Q}} \|_2^2}{c_4 w_3} \| \bm\Lambda^\iterN - \bm\Lambda^\iterO \|_2^2 + \frac{L_{\mathcal{L}}^2}{c_4(1-w_3)} \| \bm\Lambda^\iterN - \bm\Lambda_{2*} \|_2^2,
\end{equation}
where $c_4>0$ denotes the minimum eigenvalue of $\widetilde{\mathbf{A}} \widetilde{\mathbf{A}}^\top$.
Next, from the definition of $\mathbf{H}_2$, we get
\begin{equation}\label{P.Linear.Expansion.2}
\|\bm U^\iterN - \bm U_{2*}\|_{\mathbf{H}_2}^2 \le c_5 \| \bm\Lambda^\iterN - \bm\Lambda_{2*} \|_2^2 + \rho^{-1} \|\bm T^\iterN - \bm T_* \|_2^2,
\end{equation}
where $c_5>0$ denotes the maximum eigenvalue of $\widetilde{\mathbf{Q}} + \rho \widetilde{\mathbf{A}}^\top \widetilde{\mathbf{A}}$.
Combining \eqref{P.Linear.expansion.1} and \eqref{P.Linear.Expansion.2}, it holds
\begin{equation}
\begin{aligned}
{}&{} \|\bm U^\iterN - \bm U_{2*}\|_{\mathbf{H}_2}^2\\
\le {}&{} \frac{(1- w_4) \| \widetilde{\mathbf{Q}} \|_2^2}{\rho c_4 w_3} \| \bm\Lambda^\iterN - \bm\Lambda^\iterO \|_2^2 + \left( c_5+ \frac{L_{\mathcal{L}}^2}{\rho c_4(1-w_3)} \right) \| \bm\Lambda^\iterN - \bm\Lambda_{2*} \|_2^2.
\end{aligned}
\end{equation}
Hence, there exists $c_3>0$ satisfying \eqref{P.Linear.goal}, which implies the linear convergence rate.
This completes the proof of Theorem 3.4.

\end{appendix}

\begin{funding}
The research was supported in part by the National Natural Science Foundation of China (General program 12271510, General program 11871460 and program for Innovative Research Group 61621003), a grant from the Key Lab of Random Complex Structure and Data Science, CAS.
\end{funding}

\bibliographystyle{imsart-nameyear} 
\bibliography{NEL}       

\end{document}